\documentclass[a4paper,11pt]{article}
\pdfoutput=1 

\usepackage{jheppub} 

\usepackage[T1]{fontenc} 

\usepackage{slashed,cases,comment,bm,here,colortbl,amsmath,xcolor}

\usepackage{physics}
\usepackage{mathtools}
\usepackage{verbatim}
\usepackage{float}

\usepackage{graphicx}
\usepackage{multirow}
\usepackage{tabularx}
\usepackage{makecell}
\usepackage{booktabs}
\usepackage{wrapfig}

\usepackage{framed}
\definecolor{shadecolor}{rgb}{0.95,0.95,0.95}
\newenvironment{claim}{\begin{shaded}\noindent\ignorespaces}{\end{shaded}}

\title{\boldmath Dark Matter bound-state formation at higher order: a non-equilibrium quantum field theory approach}

\author[a]{Tobias Binder}

\affiliation[a]{Kavli IPMU (WPI),\\ UTIAS, The University of Tokyo, Kashiwa, Chiba 277-8583, Japan}

\author[b]{Burkhard Blobel}

\affiliation[b]{Mathematisches Institut,\\ Bunsenstr. 3-5, 37073 G\"ottingen, Germany}

\author[c]{Julia Harz}

\affiliation[c]{Physik  Department  T70,  James-Franck-Stra{\ss}e,  Technische  Universit\"at  M\"unchen,  85748  Garching, Germany}

\author[d]{Kyohei Mukaida}

\affiliation[d]{Deutsches Elektronen-Synchrotron (DESY), Notkestra{\ss}e 85, Hamburg, D-22607 Germany}

\emailAdd{tobias.binder@ipmu.jp}
\emailAdd{Burkhard.Blobel@mathematik.uni-goettingen.de}
\emailAdd{julia.harz@tum.de}
\emailAdd{kyohei.mukaida@desy.de}

\preprint{DESY 20-022, IPMU20-0014}

\abstract{The formation of meta-stable dark matter bound states in coannihilating scenarios could efficiently occur through the scattering with a variety of Standard Model bath particles, where light  bosons during the electroweak cross over or even massless photons and gluons are exchanged in the t-channel. The amplitudes for those higher-order processes, however, are divergent in the collinear direction of the in- and out-going bath particles if the mediator is massless. To address the issue of collinear divergences, we derive the bound-state formation collision term in the framework of non-equilibrium quantum field theory. The main result is an expression for a more general  cross section, which allows to compute higher-order bound-state formation processes inside the primordial plasma background in a comprehensive manner. Based on this result, we show that next-to-leading order contributions, including the bath-particle scattering, are i) collinear finite and ii) generically dominate over the on-shell emission for temperatures larger than the absolute value of the binding energy. Based on a simplified model, we demonstrate that the impact of these new effects on the thermal relic abundance is significant enough to make it worthwhile to study more realistic coannihilation scenarios.}

\begin{document} 
\maketitle
\flushbottom

\section{Introduction}
\label{sec:intro}
One of the leading dark matter (DM) candidates are Weakly Interacting Massive Particles (WIMPs)~\cite{Lee:1977ua, Ellis:1983ew, Arcadi:2017kky}, which can account for all the present DM energy fraction~\cite{Ade:2015xua} through their thermal production mechanism in the early Universe. While current upper bounds on the coupling strength to Standard Model (SM) particles put a variety of thermal dark matter candidates around the electroweak scale under tension, the TeV mass region and above remains less constrained and an attractive possibility. In such a high mass region, however, even the heaviest gauge bosons of the SM start to act as a long-range force between annihilating WIMP pairs, leading to a variety of quantum mechanical phenomena. These are important to include for predicting i) the thermal relic abundance precisely and ii) the flux of final state SM particles produced from DM annihilation in, e.g., the galactic center.

One famous example in the Minimal Supersymmetric extension of the SM (MSSM) is the traditional case of wino-like neutralino, for which the important role of quantum mechanical effects was pointed out in seminal works~\cite{Hisano:2002fk,Hisano:2003ec,Hisano:2004ds,Hisano:2006nn}. Already at the TeV mass region, the ladder exchange of the most massive SM gauge bosons enhances the probability that a slowly moving wino pair annihilates. This quantum mechanical effect is called \emph{Sommerfeld enhancement} (\textbf{SE}) \cite{doi:10.1002/andp.19314030302} or \emph{Sakharov enhancement} \cite{Sakharov:1948yq} and lowers the predicted abundance by about 50 \% compared to a tree-level computation in this case~\cite{Hisano:2006nn}. In turn, the enhancement in the cross section allows for a 30\% larger wino mass to compensate for the effect. The flux of SM particles from the present neutralino annihilation in, e.g., the galactic center is sensitive in particular to the predicted mass, since small variations can lead to Sommerfeld resonances~\cite{Hisano:2003ec}. This example shows that it is required to predict the DM mass precisely once quantum mechanical effects can drastically change the observational outcome, in order to 
constrain the model or to estimate the required exposure time for fully testing the thermal case. Various follow-up works extended the studies of the Sommerfeld effect in the MSSM to more general cases, see, e.g., refs.~\cite{Freitas:2007sa,Cirelli:2007xd,Berger:2008ti,Drees:2009gt,Hryczuk:2010zi,Hryczuk:2011tq,Hryczuk:2011vi,Beneke:2012tg,Harigaya:2014dwa,Beneke:2014hja,Beneke:2016ync,Beneke:2016jpw,ElHedri:2016onc} and~\cite{Slatyer:2009vg,Beneke:2014gja,Blum:2016nrz,Braaten:2017gpq,Braaten:2017kci,Braaten:2017dwq,Tang:2018viw} for formal aspects. For recently refined calculations of the present day SM flux in wino and higgsino models, see refs.~\cite{Beneke:2018ssm,Beneke:2019vhz,Beneke:2019gtg}.

Additional quantum mechanical effects caused by attractive long-range interactions are the existence of meta-stable bound states in the two-particle spectrum of WIMPs. Their formation and subsequent decay into SM particles can be seen as an additional channel depleting the relic density~\cite{vonHarling:2014kha} and therefore allows for even larger dark matter masses. On the one hand, bound-state effects  turn out to be negligibly small~\cite{Mitridate:2017izz} in the traditional wino case within the current description of bound-state formation (\textbf{BSF}) via the emission of an on-shell mediator. On the other hand, large corrections to the predicted mass were found in other electroweak coannihilation scenarios, e.g., the quintuplet case in the context of minimal dark matter~\cite{Mitridate:2017izz}. Similar strong effects were identified for coannihilation scenarios with colored charged particles~\cite{Ellis:2015vna,Liew:2016hqo,Fukuda:2018ufg,Harz:2018csl}. In addition to the electroweak gauge boson, photon, and gluon induced bound states, also the Higgs boson~\cite{Harz:2017dlj,Harz:2019rro} can attractively contribute to confine DM into a meta-stable bound state. Self-interacting DM~\cite{Tulin:2017ara} with new light mediators~\cite{Buckley:2009in,Aarssen:2012fx,Tulin:2013teo,Bringmann:2016din,An:2016gad,Baldes:2017gzw,Baldes:2017gzu,Kamada:2018zxi,Kamada:2018kmi,Matsumoto:2018acr,Kamada:2019jch,Ko:2019wxq}, motivated from a bottom-up approach to alleviate the diversity problem in galactic rotation curves~\cite{Kamada:2016euw,Kaplinghat:2019dhn} or in certain cases even simultaneously ameliorate the Hubble tension~\cite{Binder:2017lkj,Bringmann:2018jpr}, are further examples where long-range interactions can affect the dark matter relic abundance.

So far, the formation of dark matter bound states via the emission of an on-shell mediator was considered as the dominant process. However, it was pointed out recently that this does not necessarily have to be the dominant BSF channel during the thermal dark matter freeze-out~\cite{Binder:2019erp}. The conversion between scattering and bound states in the radiation dominated epoch can also efficiently take place via bath-particle scattering, where a mediator is exchanged virtually between a DM two-body pair and a relativistic primordial plasma particle.

\begin{figure}
\centering
\includegraphics[scale=0.80]{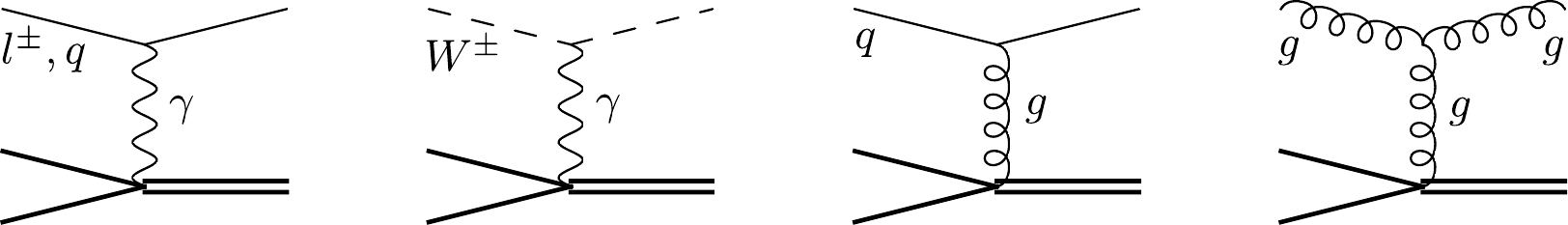}
\caption{Graphic shows examples of bound-state formation processes via bath-particle scattering. The amplitudes for these processes are divergent in the forward scattering direction of the bath particles. Parallel black solid lines represent the bound state, while open lines correspond to the initial two-particle scattering state.}
\label{fig:introbsf}
\end{figure}

While in ref.~\cite{Binder:2019erp}, a mediator with a mass larger than the binding energy was investigated, we consider massless or lighter mediators in the present work. The masses of SM force-carries is temperature dependent during the electroweak cross over, which motivates to investigate these cases more carefully.
The massless case is also relevant in coannihilation scenarios with electroweakly charged or colored particles. Examples with photons or gluons are illustrated diagrammatically in fig.~\ref{fig:introbsf}, where bound states are formed via SM bath-particle scattering. Since they all come with a large multiplicity, one may ask an interesting question by how much these additional processes contribute to the previously studied case of BSF via the emission of an on-shell mediator only.

The problem is that the bound-state formation amplitudes diverge for massless mediators in the forward scattering direction of the in- and out-going bath particles in fig.~\ref{fig:introbsf}  (see also ref.~\cite{Binder:2019erp}). Since these processes are temperature dependent, the Kinoshita--Lee--Nauenberg (KLN) theorem~\cite{Kinoshita:1962ur,Lee:1964is} is not directly applicable.
Therefore this problem can not be addressed through the computation of higher-order amplitudes in the collision term of the Boltzmann equation in zero-temperature quantum field theory.
Throughout this paper, we refer to this approach as the ``conventional Boltzmann formalism''.
Thus, the question by how much these processes could additionally deplete the dark matter relic abundance can not be answered given the conventional methods.
Naively, one might insert a Debye screening mass as a regulator. However, even this procedure is not justified for all temperatures. The energy flowing inside the mediator propagator 
from the inelastic BSF process is at least as large as the binding energy. Thus, hard thermal loop (HTL) effective field theory~\cite{Braaten:1989mz} would break down for temperatures about the binding energy or below.

The main purpose of this work is to develop a more general method of how to calculate dark matter bound-state formation processes inside the Early Universe plasma. One main focus is to address the problem of collinear divergences once massless mediators are involved and to refine the description for light mediators during the electroweak cross over. We manage to derive the BSF collision term in a more general framework of non-equilibrium quantum field theory, explicitly demonstrated for a vector mediator model in section~\ref{sec:eom}. A more general BSF cross section is defined from the collision term, expressed in terms of thermal correlation functions. Based on this novel result, we demonstrate in section~\ref{sec:LO} that our BSF cross section correctly contains the on-shell mediator emission at the perturbation order of free correlation functions. In section~\ref{sec:nlo}, a full next-to-leading order computation for the direct capture into the ground state is presented, cancellation of forward scattering divergences for bath-particle scattering and the off-shell decay is proven in a model independent way, as well as the renormalization of the ultraviolet vacuum divergence are discussed. The impact on the thermal relic density from all these new  next-to-leading order effects is shown in section~\ref{sec:comparison} and our findings discussed in section~\ref{sec:dis}. We concluded in section~\ref{sec:con}.

\section{Generalized bound-state formation cross section}
\label{sec:eom}
In order to study the infrared divergence structure of higher order bound-state formation processes within the primordial plasma environment for the case of a massless mediator, we consider for simplicity a QED-like model:
\begin{align}
\mathcal{L} \supset -g \bar{\chi} \gamma^{\mu} \chi A_{\mu} +\mathcal{L}_{\text{env}}, \label{eq:darkQED}
\end{align}
where $\chi$ is a heavy Dirac fermion, (e.g.\ coannihilating chargino partners) and the Lagrange density of the plasma environment  contains all SM fields, e.g.\ the interaction with light fermions $\mathcal{L}_{\text{env.}} \supset  -g \bar{e}\gamma^{\mu} e A_{\mu}$. The final form of the derived cross section in this section will be independent of the underlying particle content in the environment and therefore $\mathcal{L}_{\text{env}}$ can be chosen later. Temperatures around the typical chemical decoupling and below are considered, where the $\chi$ fields can be assumed to be non-relativistic to a good approximation. 
We commence from the potential non-relativistic (pNR) effective theory~\cite{Pineda:1997bj}, which allows for naturally writing the two-body bound and scattering states in terms of wave function fields.
In this framework, the transition between different two-body states are described by an ``electric'' dipole operator $\mathbf{r} \cdot \mathbf{E}(\mathbf{x},t)$, where $\mathbf{E}(\mathbf{x},t)= - \bm{\nabla} A^0(x) - \partial_t \mathbf{A} (x)$ in terms of the gauge field. In the Hamiltonian formalism, those dipole (dip) interactions are given by (see, e.g., ref.~\cite{Pineda:1997bj}):\footnote{At this stage we would already like to remark that scattering and bound states in Eq.~\eqref{eq:twobodyexp} are assumed to be separable, which is however not valid once the bound-state energy spectrum starts to overlap with the continuum in the high temperature regime for large thermal widths. This phenomena depends on what is contained in $\mathcal{L}_{\text{env}}$, see discussion in Sec.~\ref{sec:dis} and Fig.~\ref{fig:overview} for more details about the limitation of our formalism.}
\begin{align}
H_\text{dip}(t) &= - \sum_{\text{Spin}} \int \text{d}^3 x \text{d}^3 r \;  O^{\dagger}_{sr} (\mathbf{x},\mathbf{r},t) \left[ g \; \mathbf{r} \cdot \mathbf{E}(\mathbf{x},t) \right] O_{sr}(\mathbf{x},\mathbf{r},t),  \text{ where} \label{eq:dipolop} \\
O_{sr}(\mathbf{x},\mathbf{r},t) & = \nonumber \\ 
\int  \frac{\text{d}^3 K}{(2\pi)^3}  \bigg\{ & \sum_{\mathcal{B}}  e^{ - i(\mathcal{E}_{\mathcal{B}} t-\mathbf{K} \cdot \mathbf{x})} \psi_{\mathcal{B}}(\mathbf{r}) \hat c_{\mathcal{B},\mathbf{K}}^{s r} + \int \frac{\text{d}^3 k}{(2\pi)^3} e^{ - i(\mathcal{E}_{\mathbf{k}}t-\mathbf{K} \cdot \mathbf{x})} \psi_\mathbf{k}(\mathbf{r}) \hat a_{\mathbf{K}/2+\mathbf{k}}^{s} \hat b_{\mathbf{K}/2-\mathbf{k}}^{r} \bigg\}. \label{eq:twobodyexp} 
\end{align}
The dipole interaction Hamiltonian contains all possible types of two-body conversion processes, i.e., scattering-scattering, bound-bound, and scattering-bound state transitions. The general two-body field operator $O_{sr} (\mathbf{x}, \mathbf{r}, t)$ consists of the whole energy spectrum of $\chi$-field pairs, whose individual components are converted into each other through dipole interactions. Spin configurations are conserved in the transitions, represented by $s$ and $r$ for particle and anti-particle, respectively, and summed over. The components with negative energy eigenvalues $\mathcal{E}_{\mathcal{B}}$ (binding energy) are created by fundamental bound-state operators $\hat c^{\dagger}$ with quantum number $\mathcal{B}= \{ nlm \}$. The positive energy spectrum with kinetic energy $\mathcal{E}_{\mathbf{k}} = \mathbf{k}^2/(2\mu) = \mu v_{\text{rel}}^2/2$ and reduced mass $\mu$ is created by the particle operator $\hat a^{\dagger}$ and anti-particle operator $\hat b^{\dagger}$. The expansion coefficients involve the bound state $\psi_\mathcal{B}(\mathbf{r})$ and scattering state $\psi_\mathbf{k}(\mathbf{r})$ wave functions, which are the solution of the time-independent Schr\"odinger equation with a static Coulomb potential. The wave function dependence on the spin is neglected, and $\mathbf{r}$, $\mathbf{x}$ denote relative and center-of-mass coordinates, respectively.

In the following, the time evolution of the particle number density is derived in the density matrix formalism, relating the collision term to the dipole Hamiltonian.
For the sake of completeness, we first clarify the underlying assumptions.
We take an unperturbed Hamiltonian so that the number densities for the scattering and bound states are conserved and compute the BSF and dissociation rates perturbatively.
Let $\hat H$, $\hat H_{0}$, and $\hat H_{\text{int}}$ being the full, free, and interaction Hamiltonian, namely $\hat H = \hat H_{0} + \hat H_{\text{int}}$.
The unperturbed Hamiltonian is given by the summation of that for the scattering state, the bound state, and the thermal environment, $\hat H_{0} = \hat H_{\mathcal{S}} + \hat H_{\mathcal{B}} + \hat H_{\text{env}}$.
We take the initial density matrix which is factorized into a tensor product of the scattering state, the bound state, and the thermal environment: $\hat \rho (t = 0) = \hat \rho_{\mathcal{S}} \otimes \hat \rho_{\mathcal{B}} \otimes \hat \rho_{\text{env}}$.
By definition, it commutes with the unperturbed Hamiltonian: $[\hat \rho (t = 0), \hat H_{0}] = 0$.

Our starting point is the von Neumann equation in the interaction picture: $i \dot{\hat{\rho}}_{I} (t) = [ \hat H_{I} (t), \hat \rho_{I} (t) ]$,
where the density matrix and interaction Hamiltonian in the interaction picture are given by $\hat \rho_{I} = e^{i \hat H_{0} t} \hat \rho e^{- i \hat H_{0} t}$ and $\hat H_{I} = e^{i \hat H _{0} t} \hat H_{\text{int}} e^{- i \hat H_{0} t}$ respectively.
The time evolution of the particle number density is obtained from the expectation value of $\hat n_{\mathbf{k}_{\chi}} \equiv \sum_{s} \hat a_{\mathbf{k}_{\chi}}^{s \dag} \hat a_{\mathbf{k}_{\chi}}^{s}$ as follows:
\begin{align}
	\dot n_{\chi} &= \frac{1}{\operatorname{vol} (\mathbb{R}^{3})} \int \frac{\dd^{3} k_{\chi}}{(2 \pi)^{3}} \Tr \left[ \dot{\hat{\rho}}_{I} \hat n_{\mathbf{k}_{\chi}} \right] \nonumber \\
	& \simeq - \frac{1}{\operatorname{vol} (\mathbb{R}^{3})} \int \frac{\dd^{3} k_{\chi}}{(2 \pi)^{3}} \lim_{t \to \infty} \int^{t}_{0} \dd t'
	\Tr \left\{ \left[ \left[ \hat n_{\mathbf{k}_{\chi}}, \hat H_{I}(t) \right], \hat H_{I}(t')  \right] \hat \rho (t = 0) \right\}.
	\label{eq:vanNeumann}
\end{align}
In the second line, we perform the time dependent perturbation with respect $H_{I}$ and take the leading order.\footnote{
	Here only the leading order term in the interaction Hamiltonian $\hat H_I$ is kept.
	If one is interested in processes involving photons more than one, such as double photon emission, one has to consider high order terms in $\hat H_{I}$.
}
We also take $t \to \infty$ by assuming that a typical time scale of our interest is much slow compared to that in the thermal plasma.
The time evolution of the anti-particle and bound-state number density can be expressed in the same way.
Notice that the expectation value in the right-hand side of Eq.~\eqref{eq:vanNeumann} should be taken by the initial factorized density matrix.

To derive the change of the particle number density under dipole transitions, we have to evaluate the double commutator in Eq.~(\ref{eq:vanNeumann}) for the dipole interaction Hamiltonian in Eq.~(\ref{eq:dipolop}). While the dipole interaction Hamiltonian contains all possible types of two-body conversion processes, only the conversion of a scattering state into a bound state and the reverse process can change the particle number density $n_{\chi}$. Thus, it is sufficient to consider only the mixed contributions containing the product $c^{\dagger} a b$ and its hermitian conjugate part in Eq.~(\ref{eq:dipolop}). These two contribution lead to bound-state formation and the reverse process called dissociation. The computational details of the double commutator are discussed in Appendix~\ref{app:doublecommutator}.

Here we highlight the most important part in the computation in Appendix~\ref{app:doublecommutator}.
Since the dipole Hamiltonian contains the electric field, the computation of the double commutator in Eq.~\eqref{eq:vanNeumann} involves photon two-point functions.
As already emphasized, this commutator should be evaluated by means of the factorized initial density matrix, i.e., $\hat \rho (t = 0) = \hat \rho_{\mathcal{S}} \otimes \hat \rho_{\mathcal{B}} \otimes \hat \rho_{\text{env}}$ with $\hat \rho_{\text{env}} \propto e^{-\hat H_{\text{env}}/T}$.
Therefore the photon two-point functions appearing in the computation are nothing but the thermal propagators: $D^{-+}_{\mu\nu}(x-y) \equiv \langle A_\mu (x) A_\nu (y) \rangle$ for BSF and $D^{+-}_{\mu\nu}(x-y) \equiv \langle  A_\nu (y) A_\mu (x) \rangle$ for dissociation.
As well known, thermal propagators fulfill the Kubo-Martin-Schwinger relation~\cite{Kubo:1957mj,Martin:1959jp} (see also chapter "Real-time formalism prerequisites" in \cite{Binder:2018znk}):
\begin{align}
D^{-+}_{\mu\nu}(P) = [1+f_{\gamma}^\text{eq}(P^0) ] D^{\rho}_{\mu\nu}(P), \quad D^{+-}_{\mu\nu}(P) &= f_{\gamma}^\text{eq}(P^0) D^{\rho}_{\mu\nu}(P),\,
\end{align}
with $f_{\gamma}^\text{eq}$ being the equilibrium phase-space distribution obeying Bose-Einstein statistics.
The KMS condition relates both two-point functions to the spectral function, defined in coordinate space by $D^{\rho}_{\mu\nu}(x-y) \equiv \langle [A_{\mu}(x), A_{\nu}(y)]\rangle$.
This photon spectral function encodes all interactions of the primordial plasma environment.
We will outline how to estimate the effects from the thermal plasma perturbatively in Eq.~\eqref{eq:Dyson}.

With these relations, the change of the particle number density in Eq.~(\ref{eq:vanNeumann}) due to the dipole interactions in Eq.~(\ref{eq:dipolop}) is found to be given by:
\begin{align}
 \dot{n}_\chi  =
-g_{\chi}g_{\bar{\chi}}\sum_{\mathcal{B}}  &\int \frac{\text{d}^3 k_{\chi}}{(2 \pi)^3}\frac{\text{d}^3 k_{\bar{\chi}}}{(2 \pi)^3}\frac{\text{d}^3p}{(2 \pi)^3}  D^{\rho}_{\mu \nu}(\Delta E,\mathbf{p})\sum_{\text{Spin}} \mathcal{T}^{\mu}_{\mathbf{k},\mathcal{B}} (\Delta E, \mathbf{p}) \mathcal{T}^{\nu \star}_{\mathbf{k},\mathcal{B}}(\Delta E, \mathbf{p}) \nonumber \\
&\times\bigg\{ f_\chi(\mathbf k_{\chi}) f_{\bar{\chi}}(\mathbf{k}_{\bar{\chi}}) [1+f_{\gamma}^\text{eq}(\Delta E) ]   - f_{\mathcal{B}}(\mathbf{K} - \mathbf{p}) f_{\gamma}^\text{eq}(\Delta E) \bigg\},
\label{eq:BSFcollisionterm}
\end{align}
with $\Delta E =  \mathcal{E}_{\mathbf{k}}-\mathcal{E}_{\mathcal{B}}$, $\mathbf{k} = (\mathbf{k}_{\chi} - \mathbf{k}_{\bar{\chi}})/2$, $\mathbf{K} = \mathbf{k}_\chi + \mathbf{k}_{\bar\chi}$, and $\mathbf{p}$ being the three-momentum of the mediator with fixed $P^0=\Delta E$. The transition matrix elements $\mathcal{T}$ of scattering and bound states are proportional to the dipole overlap integrals:
\begin{align}
\mathcal{T}^{0}_{\mathbf{k},\mathcal{B}}(P^0,\mathbf{p}) &\equiv g \frac{i\delta^{ss^{\prime}}\delta^{rr^{\prime}}  }{\sqrt{g_{\chi} g_{\bar{\chi}}}} \mathbf{p}  \int \text{d}^3 r    \psi^{\star}_{\mathcal{B}}(\mathbf{r}) \mathbf{r} \psi_{\mathbf{k}}(\mathbf{r})  \label{eq:tmupnr2},\\
\mathcal{T}^{i}_{\mathbf{k},\mathcal{B}}(P^0,\mathbf{p}) &\equiv g \frac{i\delta^{ss^{\prime}}\delta^{rr^{\prime}}  }{\sqrt{g_{\chi} g_{\bar{\chi}}}}   P^0 \int \text{d}^3 r     \psi^{\star}_{\mathcal{B}}(\mathbf{r}) \mathbf{r}^i \psi_{\mathbf{k}}(\mathbf{r}) . \label{eq:tmupnr}
\end{align}
The transition element fulfills current conservation $P_{\mu}  \mathcal{T}_{\mathbf{k},\mathcal{B}}^{\mu}(P)  =0$, as a consequence the global symmetry of the model in Eq.~(\ref{eq:darkQED}).

The $\chi$ fields and the bound states are assumed to be in kinetic equilibrium and dilute. Their phase-space densities take the classical Maxwell-Boltzmann statistics and can be written as: $f_X= f^{\text{eq}}_Xn_X/n_X^{\text{eq}}(T) $. By noticing the standard non-relativistic thermal average in
Eq.~(\ref{eq:BSFcollisionterm}), a bound-state formation cross section can be defined as:
\begin{align}
\langle \sigma_{\mathcal{B}}^{\text{bsf}} v_{\text{rel}}  \rangle \equiv \frac{g_{\chi} g_{\bar{\chi}}}{n_{\chi}^{\text{eq}} n_{\bar{\chi}}^{\text{eq}}} \int  \frac{\text{d}^3 k_{\chi}}{(2 \pi)^3}\frac{\text{d}^3 k_{\bar{\chi}}}{(2 \pi)^3} \; (\sigma_{\mathcal{B}}^{\text{bsf}} v_{\text{rel}} ) f_\chi^{\text{eq}}(\mathbf k_{\chi}) f_{\bar{\chi}}^{\text{eq}}(\mathbf{k}_{\bar{\chi}}). \label{eq:thermalaverage}
\end{align}
Replacing in the l.h.s. of Eq.~(\ref{eq:BSFcollisionterm}), the time with cosmic time derivative and adding the usual DM annihilation part in the r.h.s., the particle number density equation can be brought into the standard form:
\begin{align}
\dot{n}_\chi + 3 H n_{\chi} = -\sum_{\mathcal{B}} \langle \sigma_{\mathcal{B}}^{\text{bsf}} v_{\text{rel}} \rangle \left[n_{\chi} n_{\bar{\chi}} - n_{\mathcal{B}} \frac{n_{\chi}^{\text{eq}} n_{\bar{\chi}}^{\text{eq}}}{n_{\mathcal{B}}^{\text{eq}}} \right] - \langle \sigma^{\text{an}} v_{\text{rel}} \rangle \left[ n_{\chi} n_{\bar{\chi}} -n_{\chi}^{\text{eq}} n_{\bar{\chi}}^{\text{eq}}\right] .
\label{eq:gen_Boltz}
\end{align}
Finally, we can identify the \emph{generalized bound-state formation cross section} as
\begin{figure}
\centering
\includegraphics[scale=1]{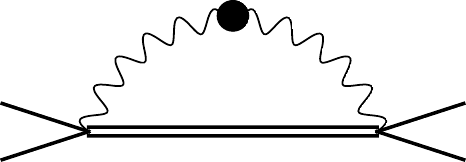}
\caption{Illustration of the generalized cross section Eq.~(\ref{eq:generalcrosssection}) in terms of a self-energy diagram, containing the interacting mediator two-point function. Parallel lines represent the bound state, while open lines correspond to the two-particle scattering state.}
\label{fig:interacting}
\end{figure}
\begin{claim}
\begin{align}
\sigma_{\mathcal{B}}^{\text{bsf}} v_{\text{rel}} &\equiv \int \frac{\text{d}^3 p}{(2\pi)^3} \left[ 1 + f_\gamma^\text{eq} (\Delta E) \right] D^{\rho}_{\mu \nu}(\Delta E,\mathbf{p}) \sum_{\text{Spin}} \mathcal{T}^{\mu}_{\mathbf{k},\mathcal{B}} (\Delta E, \mathbf{p}) \mathcal{T}^{\nu \star}_{\mathbf{k},\mathcal{B}} (\Delta E, \mathbf{p}). \label{eq:generalcrosssection}
\end{align}
\end{claim}
\noindent
This cross section is one of the central results of this work and can be diagrammatically expressed in terms of the self-energy diagram depicted in fig.~\ref{fig:interacting}. The vertices represent the leading order dipole transition contained in $\mathcal{T}_{\mathbf{k},\mathcal{B}}$. The black blob indicates the \emph{interacting spectral correlation function} $D^{\rho}$ of the mediator, which encodes all interactions with the primordial plasma environment and is the key difference compared to previous literature. It can be seen as a probability density of having a mediator excitation at a certain $\mathbf{p}$ for fixed $\Delta E$. The excitation can be on-shell but also virtually induced through the thermal environment via, e.g., bath-particle scattering. It is convenient to compute the spectral function from the retarded correlator via $D_{\mu \nu}^{\rho}= 2 \Im \left[i D^R_{\mu \nu} \right]$, since the latter obeys also in thermal field theory the Dyson-Schwinger equation, given in momentum space by
\begin{align}
D^R_{\mu \nu} = D^{R,0}_{\mu \nu} + D^{R,0}_{\mu \alpha} \Pi^{\alpha\beta}_{R} D^{R,0}_{\beta \nu} + ... \;.\label{eq:Dyson}
\end{align}
In the next section~\ref{sec:LO}, we demonstrate that BSF via on-shell mediator emission is reproduced from the free retarded propagator. In section~\ref{sec:nlo}, the first interaction term containing the retarded self-energy $\Pi_R$ is analyzed.

Although the cross section was derived for a particular model, we expect the factorization into the interacting mediator spectral function and the transition elements at the Born level to be a rather model-independent feature. In ref.~\cite{Binder:2019erp}, a direct relation between the transition matrix elements $\mathcal{T}_{\mathbf{k},\mathcal{B}}$ and relativistic amplitudes has been given. The advantage of this relation is that results of the previous literature, computing the on-shell emission only, can be used to directly determine $\mathcal{T}_{\mathbf{k},\mathcal{B}}$. For example, ref.~\cite{Harz:2018csl} provides already expressions for amplitudes for non-abelian mediators at the leading order dipole approximation, ready to be used with Eq.~(\ref{eq:generalcrosssection}). In case of a mediating gluon, this would allow to study BSF via gluon scattering (triple vertex), see the fourth diagram in fig.~\ref{fig:introbsf}, which is expected to occur at the first order of interactions in the gluon spectral function. For Yukawa interactions involving scalar mediators, one can drop the Lorentz indices in Eq.~(\ref{eq:generalcrosssection}). The corresponding transition elements, as well as many other cases, can be found in ref.~\cite{Petraki:2015hla, Petraki:2016cnz}.

\section{Recovering on-shell emission at the leading order}
\label{sec:LO}
\begin{figure}[!h]
\centering
\includegraphics[scale=0.95]{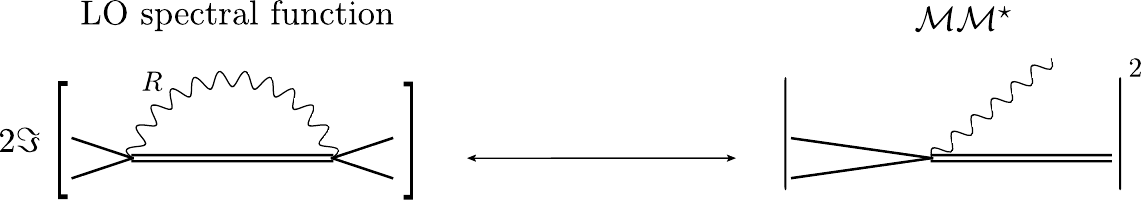}
\caption{Correspondence between the leading order self-energy contribution and the amplitude squared shown. $R$ indicates the retarded mediator propagator.}
\label{fig:LOcorresp}
\end{figure}
Consider the dominant direct capture into the ground state $nlm=100$ for the QED-like model in Eq.~(\ref{eq:darkQED}). 
The overlap integral in Eq.~(\ref{eq:tmupnr}) has been analytically computed in, e.g., ref~\cite{Petraki:2015hla}. From this result one can easily get an expression for our transition matrix elements as:
\begin{align}
\sum_{\text{spins}} \mathcal{T}^{\mu}_{\mathbf{k},100}(\Delta E,\mathbf{p})  \mathcal{T}^{\nu \star}_{\mathbf{k},100}(\Delta E,\mathbf{p}) &= \frac{4 \pi h(\zeta)}{\mu^{3} \Delta E^2} \begin{pmatrix} \mathbf{p} \cdot  \hat{\mathbf{k}} \\  \Delta E \; \hat{\mathbf{k}}   \end{pmatrix}^\mu \begin{pmatrix} \mathbf{p} \cdot  \hat{\mathbf{k}} \\  \Delta E \; \hat{\mathbf{k}}   \end{pmatrix}^\nu, \label{eq:Tgroundstate} \\
h(\zeta) &\equiv 2^6 \pi \left( \frac{2\pi \zeta}{1-e^{-2 \pi \zeta}} \right)\frac{\zeta^6}{(1+ \zeta^2)^3} e^{-4 \zeta \text{acot}\zeta},
\end{align}
with $\zeta \equiv \alpha / v_{\text{rel}}$ and $\Delta E = m_{\chi} \alpha^2(1+\zeta^{-2})/4$ for the ground-state capture. The nice feature now is that Eq.~(\ref{eq:Tgroundstate}) can be reused in Eq.~(\ref{eq:generalcrosssection}) at any order in the two-point correlation function of the photon field to determine leading and higher-order bound-state formation cross sections inside the primordial plasma environment. 

Focusing in this section on the leading order, the free retarded propagator is given by $D_{\mu \nu}^{R}(P)\big|_{\text{free}}=-i g_{\mu \nu}[P^2 + i\text{sgn}(P^0)\epsilon]^{-1}$. The imaginary part of this expression determines the free photon spectral function, which is nothing but the on-shell contribution:
\begin{align}
 D_{\mu \nu}^{\rho}(P)\big|_{\text{LO}} = - g_{\mu \nu} \text{sgn}(P^0)(2\pi)\delta(P^2).\label{eq:spectralfree}
\end{align} 

The cross section for capture into the ground state via the emission of an on-shell vector mediator is recovered (cf., e.g., \cite{Petraki:2015hla}) by inserting the free spectral function in Eq.~(\ref{eq:spectralfree}) together with the transition matrix elements in Eq.~(\ref{eq:Tgroundstate}) into the main formula Eq.~(\ref{eq:generalcrosssection}):
\begin{align}
\sigma_{100}^{\text{LO}} v_{\text{rel}}&\equiv \int \frac{\text{d}^3 p}{(2\pi)^3} \left[1+f^{\text{eq}}_\gamma(\Delta E)\right]D^{\rho}_{\mu \nu}(\Delta E,\mathbf{p})\big|_{\text{LO}} \sum_{\text{spin}}  \mathcal{T}^{\mu}_{\mathbf{k},100}(\Delta E,\mathbf{p})  \mathcal{T}^{\nu \star}_{\mathbf{k},100}(\Delta E,\mathbf{p})\nonumber \\ 
&= \frac{4 h(\zeta)\Delta E}{3 \mu^3} \left[1+f_\gamma^{\text{eq}}(\Delta E)\right].\label{eq:mBSFxsection}
\end{align}
At leading order of the mediator spectral function, the generalized BSF cross section reduces to the result one would obtain in the Boltzmann formalism based on vacuum amplitudes. In this example, the result is the same as for SM neutral hydrogen recombination, with the emission of a photon. The correspondence between the leading order self-energy diagram and the amplitudes in the vacuum field theory is shown in fig.~\ref{fig:LOcorresp}. In the regime where $T \ll \Delta E$, $\zeta \gg 1 $, and using $\lim_{\zeta \rightarrow \infty} \zeta \text{acot} \zeta = 1$, one obtains that the leading order BSF cross section Eq.~(\ref{eq:mBSFxsection}) is by approximately a factor three larger compared to the s-wave Sommerfeld enhanced annihilation cross section into two photons, see ref.~\cite{Petraki:2015hla} for details.

\section{Next-to-leading order}
\label{sec:nlo}

\begin{figure}[!h]
\centering
\includegraphics[scale=1.2]{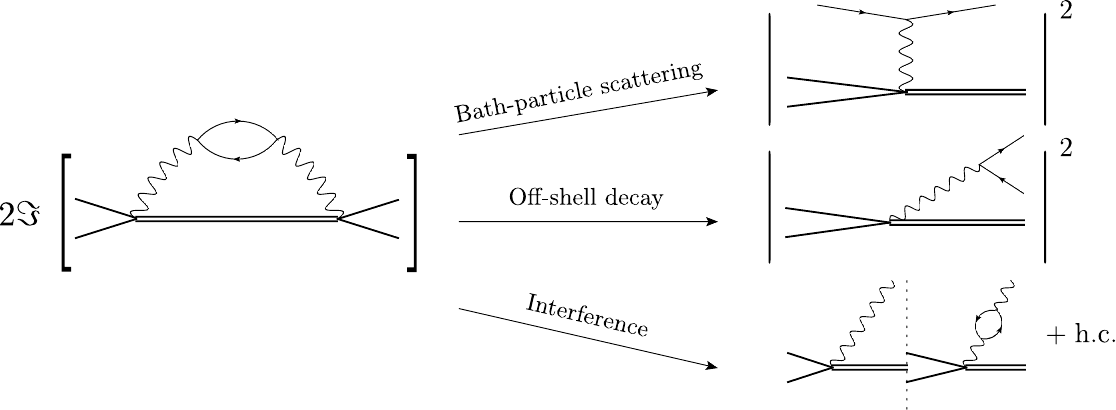}
\caption{Graphic shows the next-to-leading order contributions to the photon spectral function.}
\label{fig:NLO}
\end{figure}

We turn now to BSF processes arising at the first order of interactions with the primordial plasma environment. The imaginary part of the first interaction term in the Dyson Eq.~(\ref{eq:Dyson}) of the retarded mediator correlation function defines the next-to-leading order contribution to the spectral function and encodes the first interactions with the environment:
\begin{align}
D_{\mu \nu}^{\rho}(P) \big|_{\text{NLO}} =  2\Im\left[i D_{\mu \alpha}^{R}(P) \Pi^{\alpha \beta}_R(P) D_{\beta \nu}^{R}(P) \right].\label{eq:NLOspectralfunction}
\end{align}
Hereby, the self-energy $ \Pi^{\alpha \beta}_R(P)$ depends on the specific model. Under the assumption of a SM photon as mediator, its self-energy is a sum over various contributions with $W$-bosons, quarks and charged leptons running in the thermal loop. This is equivalent with considering all processes interacting with the plasma at next-to-leading order, for instance BSF via bath-particle scattering and off-shell mediator decay into a bath-particle pair is intrinsically taken into account in the thermal self-energy.

In the following, we concentrate on the interactions with ultra-relativistic Dirac fermions $\psi$ in the primordial plasma such that we identify $\mathcal{L}_{\text{env}} =  -g \bar{\psi}\gamma^{\mu} \psi A_{\mu}$, which resembles the interaction with SM quarks and charged leptons. Hence, the retarded self-energy contains a fermion loop as illustrated by the left diagram in fig.~\ref{fig:NLO}. We derive the retarded self-energy directly from the Wightman functions for massless fermions as described Appendix~\ref{app:retself}.  By inserting the obtained expression for the retarded photon self energy 
\begin{align}
\Pi_{\mu \nu}^R(P)&= g^2 \int \frac{\text{d}^3 k_1}{(2\pi)^3 2 |\mathbf{k}_1|} \int \frac{\text{d}^3 k_2}{(2\pi)^3 2 |\mathbf{k}_2|} \Tr[\gamma_{\mu} \slashed{K}_1 \gamma_{\nu} \slashed{K}_2] (2\pi)^3\times \nonumber  \\
&\bigg\{ \left[1-f_{\psi}^{\text{eq}}(|\mathbf{k}_1|) - f_{\psi}^{\text{eq}}(|\mathbf{k}_2|) \right] \left[ \frac{i \delta^3(\mathbf{p}+\mathbf{k}_1+\mathbf{k}_2)}{P^0+|\mathbf{k}_1|+|\mathbf{k}_2| + i \epsilon} - \frac{i \delta^3(\mathbf{p}-\mathbf{k}_1-\mathbf{k}_2)}{P^0-|\mathbf{k}_1|-|\mathbf{k}_2| + i \epsilon} \right] \nonumber \\
&+ \left[ f_{\psi}^{\text{eq}}(|\mathbf{k}_2|) - f_{\psi}^{\text{eq}}(|\mathbf{k}_1|) \right] \left[  \frac{i \delta^3(\mathbf{p}+\mathbf{k}_1-\mathbf{k}_2)}{P^0+|\mathbf{k}_1|-|\mathbf{k}_2| + i \epsilon}- \frac{i \delta^3(\mathbf{p}-\mathbf{k}_1+\mathbf{k}_2)}{P^0-|\mathbf{k}_1|+|\mathbf{k}_2| + i \epsilon} \right] \bigg\}
\end{align}
into the spectral function Eq.~(\ref{eq:NLOspectralfunction}), and the latter into the general expression for the BSF cross section Eq.~(\ref{eq:generalcrosssection}), we obtain the following expression:
\begin{align}
\sigma^{\text{NLO}}_{\mathcal{B}} v_{\text{rel}}   \equiv   \int \frac{\text{d}^3 p}{(2\pi)^3} \left[1+f^{\text{eq}}_\gamma(\Delta E)\right] &D^{\rho}_{\mu \nu}(\Delta E,\mathbf{p})\big|_{\text{NLO}}  \mathcal{T}^{\mu}_{\mathbf{k},\mathcal{B}}(\Delta E,\mathbf{p})  \mathcal{T}^{\nu \star}_{\mathbf{k},\mathcal{B}}(\Delta E,\mathbf{p})\label{eq:NLO}\\
=\int \frac{\text{d}^3 p}{(2\pi)^3} \left[1+f^{\text{eq}}_\gamma(\Delta E)\right] & \sum_{\text{spins}} \mathcal{T}^{\mu}_{\mathbf{k},\mathcal{B}}(\Delta E,\mathbf{p}) \mathcal{T}^{\nu \star}_{\mathbf{k},\mathcal{B}}(\Delta E,\mathbf{p}) \times \nonumber \\
 g^2 2\Im \left[ \left( \frac{-i}{(\Delta E+ i \epsilon )^2 - |\mathbf{p}|^2} \right)^2 \right. & \left.\int \frac{\text{d}^3 k_1}{(2\pi)^3 2 |\mathbf{k}_1|} \int \frac{\text{d}^3 k_2}{(2\pi)^3 2 |\mathbf{k}_2|} \Tr[\gamma_{\mu} \slashed{K}_1 \gamma_{\nu} \slashed{K}_2] \times \nonumber \right. \\
 \bigg\{ & \left. \frac{ \slashed\delta^3(\mathbf{p}+\mathbf{k}_1-\mathbf{k}_2)}{\Delta E +|\mathbf{k}_1|-|\mathbf{k}_2| + i \epsilon} \left[ f^{\text{eq}}_\psi(|\mathbf{k}_1|) - f^{\text{eq}}_\psi(|\mathbf{k}_2|) \right]  + \right. \nonumber \\
&\left. \frac{\slashed\delta^3(\mathbf{p}-\mathbf{k}_1+\mathbf{k}_2)}{\Delta E -|\mathbf{k}_1|+|\mathbf{k}_2| + i \epsilon} \left[ f^{\text{eq}}_\psi(|\mathbf{k}_2|) - f^{\text{eq}}_\psi(|\mathbf{k}_1|) \right]   + \nonumber \right. \\
&\left. \frac{ \slashed\delta^3(\mathbf{p}-\mathbf{k}_1-\mathbf{k}_2)}{\Delta E -|\mathbf{k}_1|-|\mathbf{k}_2| + i \epsilon} \left[1-f^{\text{eq}}_\psi(|\mathbf{k}_1|) - f^{\text{eq}}_\psi(|\mathbf{k}_2|) \right]  + \right. \nonumber \\
&\left.\frac{ \slashed\delta^3(\mathbf{p}+\mathbf{k}_1+\mathbf{k}_2)}{\Delta E +|\mathbf{k}_1|+|\mathbf{k}_2| + i \epsilon}\left[-1+f^{\text{eq}}_\psi(|\mathbf{k}_1|) + f^{\text{eq}}_\psi(|\mathbf{k}_2|) \right]     \bigg\} \right]. \nonumber
\end{align}
When integrating the above equation, the imaginary part of the integral contains a product of possible double and single poles. The former originate from the squared photon propagator $D^2$ and the latter from the self-energy $\Pi_R$ in Eq.~(\ref{eq:NLOspectralfunction}).

Taking the imaginary part of single poles corresponds to putting the fermions in the loop as on-shell, while the photons remain virtual. From top to bottom, the four single poles belong to: BSF via particle\footnote{To identify the bath-particle scattering case, the identity $\left[1+f_\gamma^{\text{eq}}(\Delta E)\right]\left[f_{\psi}^{\text{eq}}(|\mathbf{k}_1|)-f_{\psi}^{\text{eq}}(|\mathbf{k}_1|+\Delta E)\right]=f_{\psi}^{\text{eq}}(|\mathbf{k}_1|) \left[1-f_{\psi}^{\text{eq}}(\Delta E+|\mathbf{k}_1|)\right]$ is helpful. Similarly, all three other possibilities can be identified. We checked that the resulting cross sections from the imaginary part of the single poles only are identical to the cross sections obtained in the conventional Boltzmann formalism, and that they are collinear divergent.} and anti-particle scattering, as well as BSF via off-shell decay of the photon into a pair of bath particles and the reverse process. The reverse process of off-shell decay is an exception and does not contribute to the imaginary part for kinematical reason (Note that $\Delta E$ is always positive). All others are graphically represented by the top and middle diagram on the right-hand side of fig.~\ref{fig:NLO}.

The contributions arising from the imaginary part of the double pole correspond to the on-shell emission at next-to-leading order with a temperature dependent fermion loop. We call these contributions in the following interference terms, collectively represented by the bottom right diagram in fig.~\ref{fig:NLO}.

The total BSF cross section amounts to a sum over all imaginary parts of the aforementioned single and double poles. The bath-particle scattering and off-shell decay, which arise from the imaginary part of single poles, diverge in the case of a massless photon for zero opening angle of the bath particles ($\hat{\mathbf{k}}_1 \cdot \hat{\mathbf{k}}_2 \rightarrow 1$). We will demonstrate that this divergence is exactly canceled by the collinear divergence appearing in the interference terms. In the \emph{conventional} Boltzmann approach (i.e. considering only bath-particle scattering without on-shell emission at NLO and finite temperature fermion loop), there is no other term regulating this divergence, while in our approach the cancellation will be shown to happen automatically. The loop contains also UV divergent vacuum parts, i.e.\ $f_{\psi}$ independent terms in Eq.~(\ref{eq:NLO}), which can be renormalized with the standard counter terms.

In the following, we show in section~\ref{sec:cumputation} the main steps for computing BSF at next-to-leading order for the special case of capture into the ground state. In section~\ref{sec:cancellation}, we provide a general proof for the cancellation of the appearing collinear divergences in our formalism. The result for the next-to-leading order BSF cross section is compared to the leading order contribution (on-shell mediator emission) in section~\ref{sec:crosssections}. We set our results into context to the massive mediator case in section~\ref{sec:massive}.

\subsection{Computational results for direct capture into the ground state}
\label{sec:cumputation}
The discussion of the cancellation of the forward scattering divergences and the renormalization of the UV divergences can be separated by noticing that the self-energy can be written as a sum over finite temperature and zero temperature parts. According to this separation, it is useful to define cross sections respectively:
\begin{align}
\frac{1}{4 \pi}\int \text{d} \Omega_{\mathbf{k}} (\sigma^{\text{NLO}}_{100} v_{\text{rel}}) = (\sigma^{\text{NLO}}_{100} v_{\text{rel}})_{T=0}  + (\sigma^{\text{NLO}}_{100} v_{\text{rel}})_{T\neq 0},
\label{eq:separation_vac_nonvac}
\end{align}
where the angular averaged parts are defined as
\begin{align}
(\sigma^{\text{NLO}}_{100} v_{\text{rel}})_{T=0} \equiv \frac{1}{4 \pi}\int \text{d} \Omega_{\mathbf{k}}  \int \frac{\text{d}^3 p}{(2\pi)^3} &\left[1+f^{\text{eq}}_\gamma(\Delta E)\right] \mathcal{T}^{\mu}_{\mathbf{k},100}(\Delta E,\mathbf{p})  \mathcal{T}^{\nu \star}_{\mathbf{k},100}(\Delta E,\mathbf{p}) \times \label{eq:vaccrossdef} \\
&  2\Im\left[i D_{\mu \alpha}^{R}(\Delta E,\mathbf{p})  \Pi^{\alpha \beta}_R(\Delta E,\mathbf{p})  D_{\beta \nu}^{R}(\Delta E,\mathbf{p})  \right]_{T=0}, \nonumber \\
(\sigma^{\text{NLO}}_{100} v_{\text{rel}})_{T\neq 0} \equiv \frac{1}{4 \pi}\int \text{d} \Omega_{\mathbf{k}}  \int \frac{\text{d}^3 p}{(2\pi)^3}& \left[1+f^{\text{eq}}_\gamma(\Delta E)\right]  \mathcal{T}^{\mu}_{\mathbf{k},100}(\Delta E,\mathbf{p})  \mathcal{T}^{\nu \star}_{\mathbf{k},100}(\Delta E,\mathbf{p}) \times\\
& 2\Im\left[i D_{\mu \alpha}^{R}(\Delta E,\mathbf{p})  \Pi^{\alpha \beta}_R(\Delta E,\mathbf{p})  D_{\beta \nu}^{R}(\Delta E,\mathbf{p})  \right]_{T \neq 0}.\nonumber
\end{align}
These two cross sections are computed separately in the following. The transition elements $\mathcal{T}$ for the direct capture into the ground state are given in Eq.~(\ref{eq:Tgroundstate}). Out of computational reasons it is already reasonable to perform here the angular average over the orientation of the DM relative momentum and it naturally arises in the thermal average in Eq.~(\ref{eq:thermalaverage}). The following identities for the direct capture into the ground state will be helpful to simplify appearing expressions:
\begin{align}
&\frac{1}{4\pi}\int\text{d}\Omega_{\mathbf{k}} \sum_{\text{spins}} \mathcal{T}^{\mu}_{\mathbf{k},100}(\Delta E,\mathbf{p})  \mathcal{T}^{\nu \star}_{\mathbf{k},100}(\Delta E,\mathbf{p})(g^{\mu \nu}P^2-P^{\mu} P^{\nu} )\nonumber \\& =\frac{4 h(\zeta) \Delta E}{3 \mu^3} \frac{3 \pi}{\Delta E^3}(\Delta E^2-\mathbf{p}^2) (\mathbf{p}^2/3-\Delta E^2), \\
&\frac{1}{4\pi}\int\text{d}\Omega_{\mathbf{k}} \sum_{\text{spins}} \mathcal{T}^{\mu}_{\mathbf{k},100}(\Delta E,\mathbf{p})  \mathcal{T}^{\nu \star}_{\mathbf{k},100}(\Delta E,\mathbf{p})\Tr[\gamma_{\mu} \slashed{K}_1 \gamma_{\nu} \slashed{K}_2]=\nonumber \\
&\frac{16 \pi h(\zeta)}{3 \mu^3 \Delta E^2} \bigg\{ 2 \left(|\mathbf{k}_1||\mathbf{k}_2|\mathbf{p}^2-\Delta E |\mathbf{k}_1| \mathbf{p}\cdot \mathbf{k}_2 - \Delta E |\mathbf{k}_2| \mathbf{p}\cdot \mathbf{k}_1 +  \Delta E^2 \mathbf{k}_1\cdot \mathbf{k}_2\right) \nonumber \\ &- \left(\mathbf{p}^2-3 \Delta E^2 \right)\left( |\mathbf{k}_1||\mathbf{k}_2|-\mathbf{k}_1\cdot \mathbf{k}_2\right) \bigg\}.
\end{align}

\paragraph*{Vacuum contribution}
The renormalized retarded self-energy at zero temperature can be obtained from the conventional Euclidean time-ordered self-energy through analytic continuation. In the $\overline{\text{MS}}$-scheme, the retarded photon self-energy for vacuum polarization with approximately massless fermions running in the loop is given by:
\begin{align}
i\Pi_{\mu \nu}^R (P)\big|_{T=0} = -\left(g_{\mu \nu}P^2-P_{\mu} P_{\nu} \right) \frac{g^2}{12 \pi^2} \left[ \ln\left(\frac{P^2+i\text{sign}(P^0)\epsilon }{-\mu^2_0} \right)-\frac{5}{3} \right].
\end{align}
The logarithm has a complex contribution for time-like $P^2$, which originates from the off-shell decay into a massless bath-particle pair. Computing the integral over the imaginary parts in Eq.~(\ref{eq:vaccrossdef}), we realize that the NLO cross section factorizes
\begin{align}
(\sigma_{100}^{\text{NLO}} v_{\text{rel}})_{T = 0} = (\sigma_{100}^{\text{LO}} v_{\text{rel}}) \left[\frac{\alpha}{\pi} \lim_{\epsilon \searrow 0}  R_{\epsilon} \right],\label{eq:vacxsec}
\end{align}
into the LO contribution as given in Eq.~(\ref{eq:mBSFxsection}) and the NLO contribution with 
\begin{align}
R_{\epsilon} &= \frac{1}{\pi} \int_0^{\infty} \text{d} |\mathbf{p}| \Im \bigg\{ \frac{\mathbf{p}^2 (\Delta E^2-\mathbf{p}^2 )\left(\mathbf{p}^2-3\Delta E^2\right)}{3\Delta E^3\left[ (\Delta E+ i \epsilon)^2-\mathbf{p}^2  \right]^2}  \left[\ln\left(\frac{(\Delta E+ i \epsilon)^2-\mathbf{p}^2 }{-\mu^2_0} \right)-\frac{5}{3} \right]  \bigg\}.\label{vac:intvac}
\end{align}
Performing the integral and taking $\epsilon \rightarrow 0$, we obtain a finite result (see appendix~\ref{app:vaccontour} for the details on the contour integration):
\begin{align}
\lim_{\epsilon \searrow 0} R_{\epsilon} = \frac{1}{3} \left[\ln\left(\frac{\Delta E^2}{\mu^2_0/4} \right)- \frac{10}{3}\right].
\end{align}
These vacuum corrections are shown in fig.~\ref{fig:vac}, shared and discussed in more detail in appendix~\ref{app:scale}. From now on, the renormalization scale is fixed to the ground state binding energy $\mu_0 = E_{100}= \mu \alpha^2/2$.

As naively expected at $T=0$ from the KLN theorem, the collinear divergences in real and virtual corrections cancel each other. Hence, for a finite, physical cross section, both contributions, off-shell decays as well as the virtual correction to the on-shell emission have to be taken into account. This similarly suggests that when considering off-shell decays in a thermal plasma that we have to consider similarly on-shell emission at next-to-leading order at finite temperatures.

While a consistent treatment of BSF at NLO is non-trivial in the conventional approach (e.g. with respect to double counting of real intermediate states), our advocated formalism will take care of all crucial subtleties such as finite temperature effects, double counting of real intermediate states, and the ``automatic'' cancellation of all appearing infrared divergences, as we will discuss in the following.

\paragraph*{Finite temperature contributions}
Similar to Eq.~(\ref{eq:vacxsec}), the finite temperature part of the NLO cross section can be written as
\begin{align}
(\sigma_{100}^{\text{NLO}} v_{\text{rel}})_{T\neq 0} &= (\sigma_{100}^{\text{LO}} v_{\text{rel}}) \left[\frac{\alpha}{\pi} \lim_{\epsilon \searrow 0} \sum_{\sigma_1,\sigma_2} R^{\sigma_1\sigma_2}_{\epsilon}\right],\label{eq:ftxsec}
\end{align}
where the remaining part to compute is the dimensionless function:\footnote{In principle, it does not matter which integral is chosen to perform the integration over the poles. However, we would like to share some insights why we find this particular order especially suited. Here, we have chosen the integral over $\mathbf{k}_2$ in Eq.~(\ref{eq:NLO}) to first perform the simple integration over the momentum conserving delta functions, and then $\mathbf{p}$ to perform integration over the poles. In this order, the collinear divergence occurs at $\tau \equiv \hat{\mathbf{p}} \cdot \hat{\mathbf{k}}_1\rightarrow 1$, which is physically not anymore the zero opening angle between the bath particles but related to that divergence. In a different order where $\mathbf{p}$ integration over the delta functions is performed first, the collinear divergences occur when the opening angle of the bath particles approaches zero, i.e., $\hat{\mathbf{k}}_1 \cdot \hat{\mathbf{k}}_2 \rightarrow 1$. This order introduces coordinate singularities in addition, which one can get rid off by transforming into certain elliptical coordinates. The final result in these coordinates would be the same as in the order of integration chosen in this section, but more difficult to physically interpret. Finally, we would like to remark that one is not allowed to choose different integration orders for the single and double poles. This is because only the simultaneous integral over both contributions is in general finite,
while the individual terms can diverge. When separating the divergent integrals over single and double poles, we find that one can still show that the collinear divergences cancel at the end, but one misses important boundary terms and the final result would be different.
}
\begin{align}
R^{\sigma_1\sigma_2}_{\epsilon}&\equiv \frac{1}{\pi}  \int_0^{\infty} \text{d} |\mathbf{k}| \int_{-1}^{1} \text{d} \tau \int_0^{\infty} \text{d} |\mathbf{p}| \Im \left[G^{\sigma_1\sigma_2}_{\epsilon}(|\mathbf{p}|,\tau,|\mathbf{k}|)\right]. \label{eq:rdef}
\end{align}
In this compact notation, the summation over $\sigma_i \in \{+,-\} $ in Eq.~(\ref{eq:ftxsec}) takes into account all the finite temperature contributions contained in the self-energy in Eq.~(\ref{eq:NLO}).
To arrive here, integration over $\mathbf{k}_2$ in Eq.~(\ref{eq:NLO}) was performed over the momentum conserving delta function and $\mathbf{k}_1$ was relabeled by $\mathbf{k}$. The angular integration variable is $\tau \equiv \hat{\mathbf{p}} \cdot \hat{\mathbf{k}}$. The function $G^{\sigma_1\sigma_2}_{\epsilon}$ contains a product of the single and double poles, as given by
\begin{align}
G^{\sigma_1\sigma_2}_{\epsilon}(|\mathbf{p}|,\tau,|\mathbf{k}|) &\equiv \frac{ F^{\sigma_1\sigma_2}(|\mathbf{p}|,\tau,|\mathbf{k}|)}{\left([\Delta E+i \epsilon]^2 - \mathbf{p}^2 \right)^{2} \left(\Delta E + \sigma_1|\mathbf{k}| - \sigma_2 |\mathbf{p}+\sigma_1\mathbf{k}| + i \epsilon\right)}, \label{eq:gsigmas}
\end{align}
where for the numerator we get
\begin{align}
& F^{\sigma_1\sigma_2}(|\mathbf{p}|,\tau,|\mathbf{k}|)  \equiv \left[ \sigma_1 f^{\text{eq}}_\psi(|\mathbf{k}|) - \sigma_2 f^{\text{eq}}_\psi(|\mathbf{p}+\sigma_1 \mathbf{k}|) \right] \frac{\mathbf{k}^2\mathbf{p}^2}{\Delta E^3 |\mathbf{p}+\sigma_1 \mathbf{k}|} \times \label{eq:f12} \\
 &  \bigg\{  \left[\mathbf{p}^2-3 \Delta E^2 \right]\left[ |\mathbf{p}+\sigma_1 \mathbf{k}|-\sigma_2 (\sigma_1|\mathbf{k}|+|\mathbf{p}|\tau)\right]- \nonumber \\ & 2  \left[\mathbf{p}^2|\mathbf{p}+\sigma_1 \mathbf{k}|-\Delta E  \sigma_2 |\mathbf{p}|(|\mathbf{p}| + \sigma_1|\mathbf{k}| \tau) - \Delta E |\mathbf{p}+\sigma_1 \mathbf{k}| |\mathbf{p}|\tau +  \Delta E^2 \sigma_2 (\sigma_1|\mathbf{k}| + |\mathbf{p}|\tau)\right]  \bigg\} \nonumber.
\end{align}
For the integration over $|\mathbf{p}|$ in Eq.~(\ref{eq:rdef}), we consider the analytic continuation $|\mathbf{p}|\rightarrow z$ and perform the integration over the single and double poles of the function in Eq.~(\ref{eq:gsigmas}). In appendix~\ref{app:contourFT}, we show that due to the imaginary part in Eq.~(\ref{eq:rdef}), $F^{\sigma_1\sigma_2}$ holomorphic on the real line, and integration range from 0 to infinity, only the real and positive poles contribute. All the real and positive poles are listed in table~\ref{tab:poles}, together with their existence criteria. We denote the double pole as $z_0$ and the possible two single poles as $z_{p/m}$ for each $\sigma_1 \sigma_2$ configuration in Eq.~(\ref{eq:gsigmas}). The table can be used to simply write down the $R^{\sigma_1\sigma_2}$ functions expressed in terms of the residues as:
\begin{align}
R^{++}_{0} &= \int_0^{\infty} \text{d} |\mathbf{k}| \int_{-1}^{1} \text{d} \tau \left[ \text{Res}(G_{0}^{++},z_0) + \text{Res}(G_{0}^{++},z_p) \right], \label{eq:Rpp}\\
R^{+-}_{0} &= \int_0^{\infty} \text{d} |\mathbf{k}| \int_{-1}^{1} \text{d} \tau \left[ \text{Res}(G_{0}^{+-},z_0) \right], \label{eq:Rpm} \\
R^{-+}_{0} &= \int_0^{\Delta E /2} \text{d} |\mathbf{k}| \int_{-1}^{1} \text{d} \tau \left[ \text{Res}(G_{0}^{-+},z_0) +\text{Res}(G_{0}^{-+},z_p)\right] \label{eq:Rmp}\\
&+ \int_{\Delta E /2}^{\Delta E} \text{d} |\mathbf{k}|  \bigg\{ \int_{\sqrt{\frac{2\Delta E |\mathbf{k}|-\Delta E^2}{|\mathbf{k}|^2}}}^{1} \text{d} \tau  \left[ \text{Res}(G_{0}^{-+},z_0) +\text{Res}(G_{0}^{-+},z_p)-\text{Res}(G_{0}^{-+},z_m)\right]\nonumber  \\
& \phantom{\int_{\Delta E /2}^{\Delta E} \text{d} |\mathbf{k}| } + \int_{-1}^{\sqrt{\frac{2\Delta E |\mathbf{k}|-\Delta E^2}{|\mathbf{k}|^2}}}\text{d} \tau  \left[ \text{Res}(G_{0}^{-+},z_0)\right] \bigg\} \nonumber \\
&+\int_{\Delta E}^{\infty} \text{d} |\mathbf{k}| \int_{-1}^{1} \text{d} \tau \left[ \text{Res}(G_{0}^{-+},z_0) \right], \nonumber \\
R^{--}_{0}&= R^{++}_{0}. \label{eq:Rmm}
\end{align}
The different signs in front of the residues originate from the $i\epsilon$ in Eq.~(\ref{eq:gsigmas}), where the $z_m$ pole was constructed to lie always in the lower complex half plane for finite epsilon and the other ones in the upper half plane, see appendix~\ref{app:contourFT} for details. The last equality follows from symmetry considerations, which one can directly see from Eq.~(\ref{eq:NLO}) by interchanging the labels $\mathbf{k}_1$ and $\mathbf{k}_2$ for the corresponding term \footnote{Physically, this symmetry states that BSF via particle or anti-particle scattering is the same.}.

\begin{table*}
\begin{center}
\begin{tabular}{c||c|c|c} 
\toprule 
   $\sigma_1\sigma_2$ & $z_0$ & $z_p$& $z_m$   \\ \midrule
    $++$ & $\Delta E$ & $-|\mathbf{k}|\tau + \sqrt{\mathbf{k}^2\tau^2 + \Delta E^2 + 2 \Delta E |\mathbf{k}|}$ & $-$ \\ \midrule 
    $+-$ & $\Delta E$ & $-$ & $-$ \\  \midrule 
    $-+$ & $\Delta E$ & $|\mathbf{k}|\tau + \sqrt{\mathbf{k}^2\tau^2 + \Delta E^2 - 2 \Delta E |\mathbf{k}|}$, & $|\mathbf{k}|\tau - \sqrt{\mathbf{k}^2\tau^2 + \Delta E^2 - 2 \Delta E |\mathbf{k}|}$, \\ 
& & if $  \bigg\{ \Delta E/2 \leq |\mathbf{k}| \leq \Delta E  $ & if $\Delta E/2 \leq |\mathbf{k}| \leq \Delta E  $\\
& & $ \land \tau \geq \sqrt{\frac{2\Delta E |\mathbf{k}|-\Delta E^2}{|\mathbf{k}|^2}} \bigg\} $ & $\land \tau \geq \sqrt{\frac{2\Delta E |\mathbf{k}|-\Delta E^2}{|\mathbf{k}|^2}}.$ \\
& & or $ 0 \leq |\mathbf{k}| \leq \Delta E/2. $ &  \\ \bottomrule
\end{tabular}
\end{center}
\caption{Summary of the real and positive poles of Eq.~(\ref{eq:gsigmas}), as well as their existence criteria.}
\label{tab:poles}
\end{table*}

The list of all finite temperature contributions, from Eq.~(\ref{eq:Rpp}) to Eq.~(\ref{eq:Rmm}), is subject to further discussion. BSF via bath-particle scattering is contained in the single pole contribution $\text{Res}(G_{0}^{++},z_p)$ and the finite temperature part of the off-shell decay of the vector mediator into a bath-particle pair is contained in the $z_p$ residues in $R^{-+}$. The individual residue would diverge in the limit $\tau \rightarrow 1$, reflecting the collinear divergence. However, the double pole contribution with $z_0$ diverges in the limit as well, but with opposite sign, resulting in the fact that the sum over double and single pole contributions remains finite in collinear direction. A rather general proof for the cancellation of collinear divergences is presented in the next section. Therein, it is also explained why the terms where no single poles exist are finite. The remaining $\mathbf{k}$ and $\tau$ integrals are all finite and we show their numerically obtained values in fig.~\ref{fig:rfunction}, shared in the appendix~\ref{app:indFT}. Worthwhile to note is that the most dominant contributions are $R^{++}$ and $R^{--}$, which contain BSF via particle and anti-particle, respectively.

\subsection{Proof for the cancellation of collinear divergences}
\label{sec:cancellation}

From the mathematical point of view, collinear divergences occur since in the collinear limit $\tau \to 1$ the single pole $z_p$ approaches the double pole $z_0$ as
\begin{align}
	z_p-z_0 &= (1-\tau) \frac{\sigma_1|\mathbf{k}| \Delta E}{\Delta E + \sigma_1|\mathbf{k}|} + \mathcal{O}((1-\tau)^2). \label{eq:colllimit}
\end{align}
To make that clear, let us take a closer look on the residues of a function with a double pole at $z_0$ and a single pole at $z_p$, approaching each other in the collinear limit as, e.g., in Eq.~(\ref{eq:colllimit}). Every holomorphic function having such kind of pole structure can be written as
\begin{align}
	G(z) = \frac{H(z)}{(z-z_0)^2(z-z_p)},
	\label{eq:rewriteGH}
\end{align}
where $H$ is holomorphic at $z=z_0$ and $z=z_p$. Using this general form, the residues of $G$ are given by:
\begin{align}
	\Res(G,z_0) &= - \frac{H(z_0)}{(z_p-z_0)^2} - \frac{H'(z_0)}{z_p-z_0},									\label{eq:ResG0}\\
	\Res(G,z_p) &= \frac{H(z_p)}{(z_p-z_0)^2}																					\nonumber\\
	&= \frac{H(z_0)}{(z_p-z_0)^2} + \frac{H'(z_0)}{z_p-z_0} + \frac{1}{2}H''(z_0) + \mathcal{O}(z_p-z_0),		\label{eq:ResGp}
\end{align}
In the last line we expanded $H(z_p)$ around $z_p=z_0$.
Clearly, each individual residue is divergent in the collinear limit, when $z_p \rightarrow z_0$.
However, the colliner divergent terms from Eq.~(\ref{eq:ResG0}) and Eq.~(\ref{eq:ResGp}) occur with opposite sign and hence cancel each other in their sum.
Let us comment on the remaining term $\frac12 H''(z_0)$ whose appearance one can intuitive understand. Since the double pole merges with the single pole in the collinear limit, the function $G|_{\tau=1} = \frac{H(z)}{(z-z_0)^3}$ has a triple pole. And indeed, its residue gives the result expected from Eq.~(\ref{eq:ResG0}) and Eq.~(\ref{eq:ResGp}) as 
\begin{align}
\Res(G|_{\tau=1},z_0) &= \frac{1}{2} \lim_{z \to z_0} \frac{\text{d}^2}{\text{d} z ^2} \left[ (z-z_0)^3 \frac{H(z)}{(z-z_0)^3} \right] \nonumber\\
&=\frac{1}{2} H''(z_0) \nonumber \\
&= \lim_{\tau\to 1}\left[\Res(G,z_p)+\Res(G,z_0) \right],
\end{align}
without any collinear divergence.  This proves the collinear finiteness for \emph{any} $H$ holomorphic at $z=z_0$ and $z=z_p$ and is in particular fulfilled for any $F^{\sigma_1\sigma_2}$ in Eq.~(\ref{eq:f12}) for capture into the ground state.

From the discussion above, it follows that $R^{++}_0$ as well as $R^{--}_0$ in Eq.~(\ref{eq:Rpp}) and Eq.~(\ref{eq:Rmm}) are finite. For $\sigma_1=+$, $\sigma_2=-$, the single poles do not exists as $\Delta E + |\mathbf{k}_1| + |\mathbf{k}_2|$ is always nonzero. Hence the singularity at $z=z_{p}$ is removable and Eq.~(\ref{eq:ResG0}) as well as the singular part of Eq.~(\ref{eq:ResGp}) vanish. Therefore Eq.~(\ref{eq:Rpm}) is collinear finite. It remains to discuss Eq.~(\ref{eq:Rmp}), where $\sigma_1=-$ and $\sigma_2=+$. As one can see, the collinear divergences are canceled in the first and the second line. In the third line $\tau$ is smaller than $1$. The last line of Eq.~(\ref{eq:Rmp}) is collinear finite since there are no single poles as $\Delta E - |\mathbf{k}_1| - |\mathbf{k}_2|$ is always negative for $|\mathbf{k}_1| > \Delta E$. Analog to the case $\sigma_1=+$, $\sigma_2=-$ the singularities vanish.

While the residue of the single pole $\Res(G,z_p)$ and the residue of the double pole $\Res(G,z_0)$ diverge individually in the collinear limit, we have shown in summary that the sum of both terms remains finite. In the Boltzmann formalism only the single pole $\Res(G,z_p)$ would occur, and hence the collision term would be ill-defined. This makes the thermal field theory approach necessary for studying BSF at higher order, at least if massless gauge bosons or light mediators (for how light see section~\ref{sec:massive}) are involved. 

It remains to discuss the residue at $z_m$, only present in $R^{-+}_0$. Similarly, $\Res(G_0,z_p)$ and $\Res(G_0,z_m)$ have poles if $z_p = z_m$. This situation occurs if $|\tau| = \sqrt{\frac{2\Delta E |\mathbf{k}| - \Delta E^2}{|\mathbf{k}|^2}} =: \tau_*$ and therefore it is only present in the second line of Eq.~(\ref{eq:Rmp}). Because of the relative signs these singularities do not cancel. However, due to the choice of the order of the integration (see discussion in Sec.~\ref{sec:cumputation}, paragraph \textbf{Finite temperature contributions}) the singularities only grow like $(\tau - \tau_*)^{-\frac{1}{2}}$. For that reason the integrating over $\tau$ does not cause any problems. Finally, the last exception occurs if $z_0 = z_p = z_m$. This is the case for $\tau = 1$, $|\mathbf{k}| = \Delta E$. In this case the function $F^{-+}$ given in Eq.~(\ref{eq:f12}) goes to zero and the integral remains finite.

\subsection{Comparison of ground state capture at leading and next-to-leading order}
\label{sec:crosssections}

\begin{figure}[H]
\centering
\includegraphics[scale=0.33]{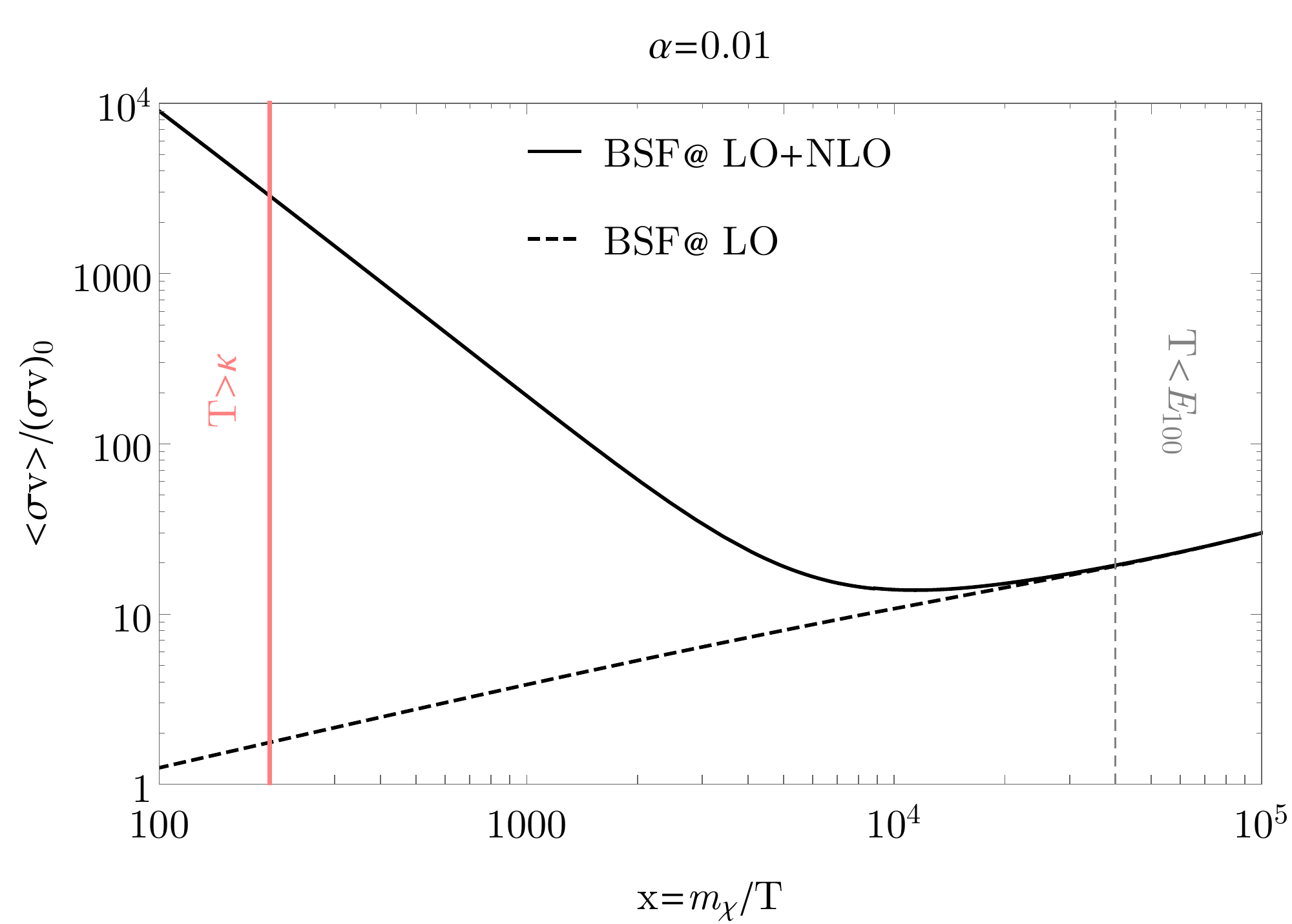}
\includegraphics[scale=0.33]{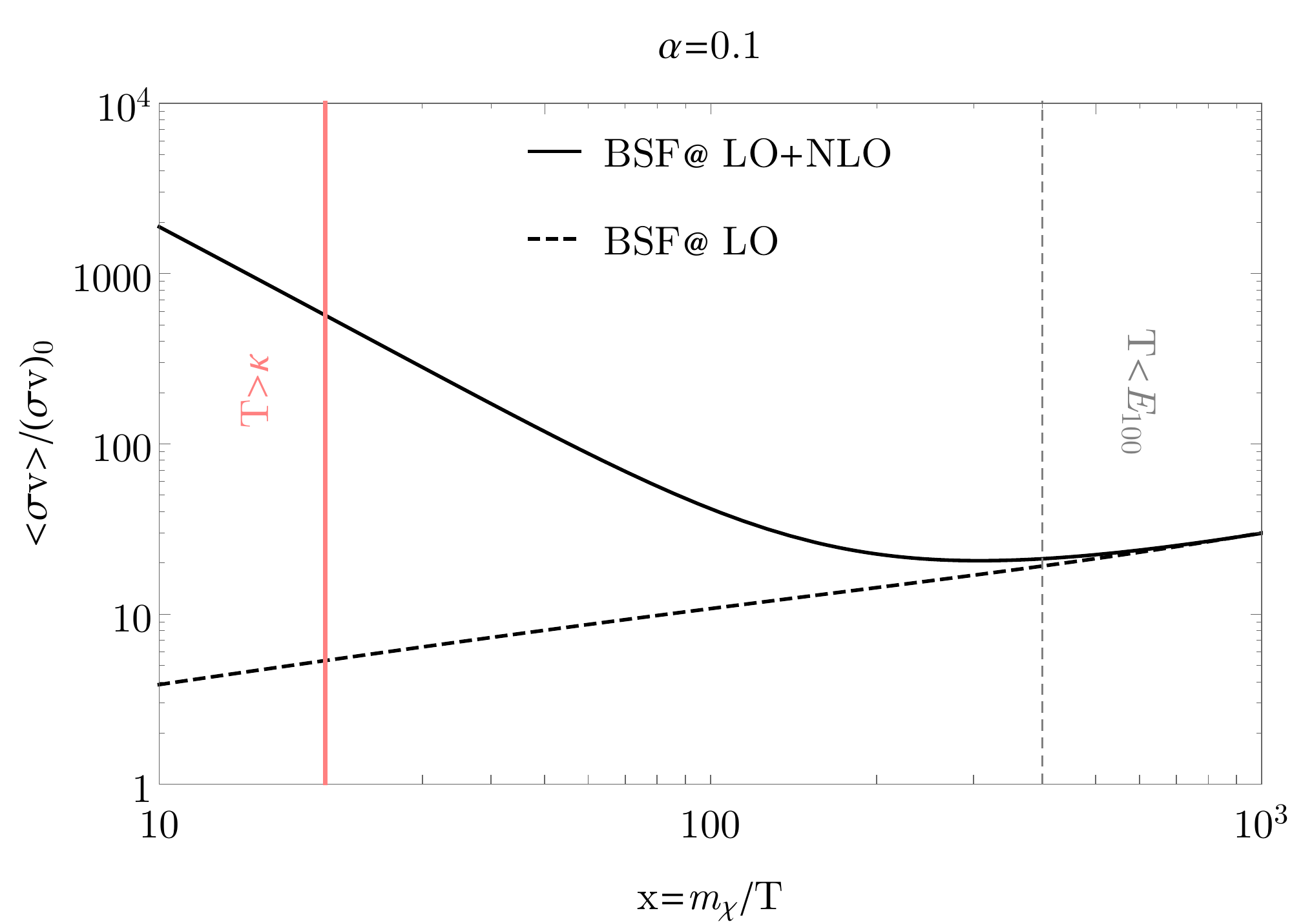}
\includegraphics[scale=0.33]{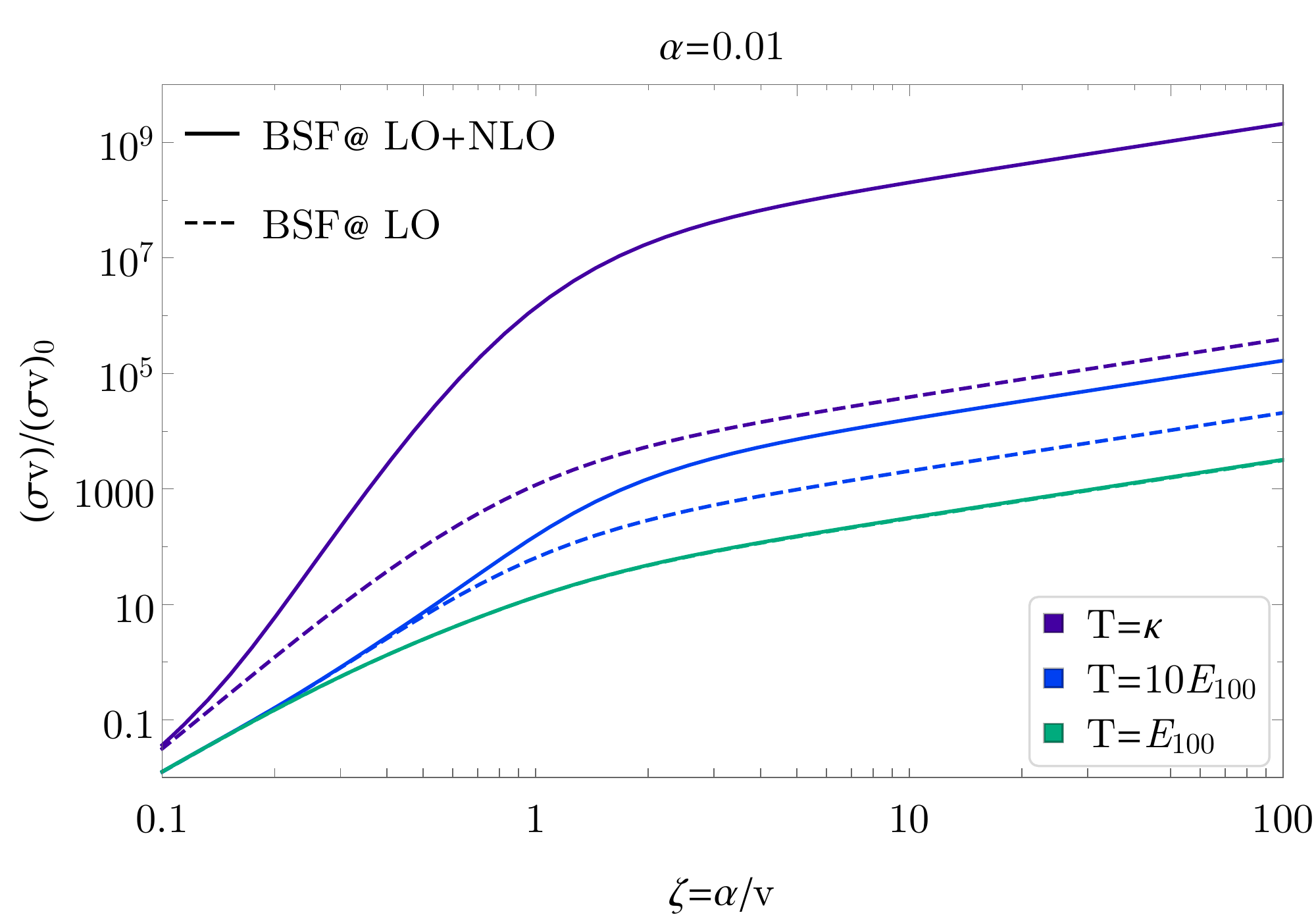}
\includegraphics[scale=0.33]{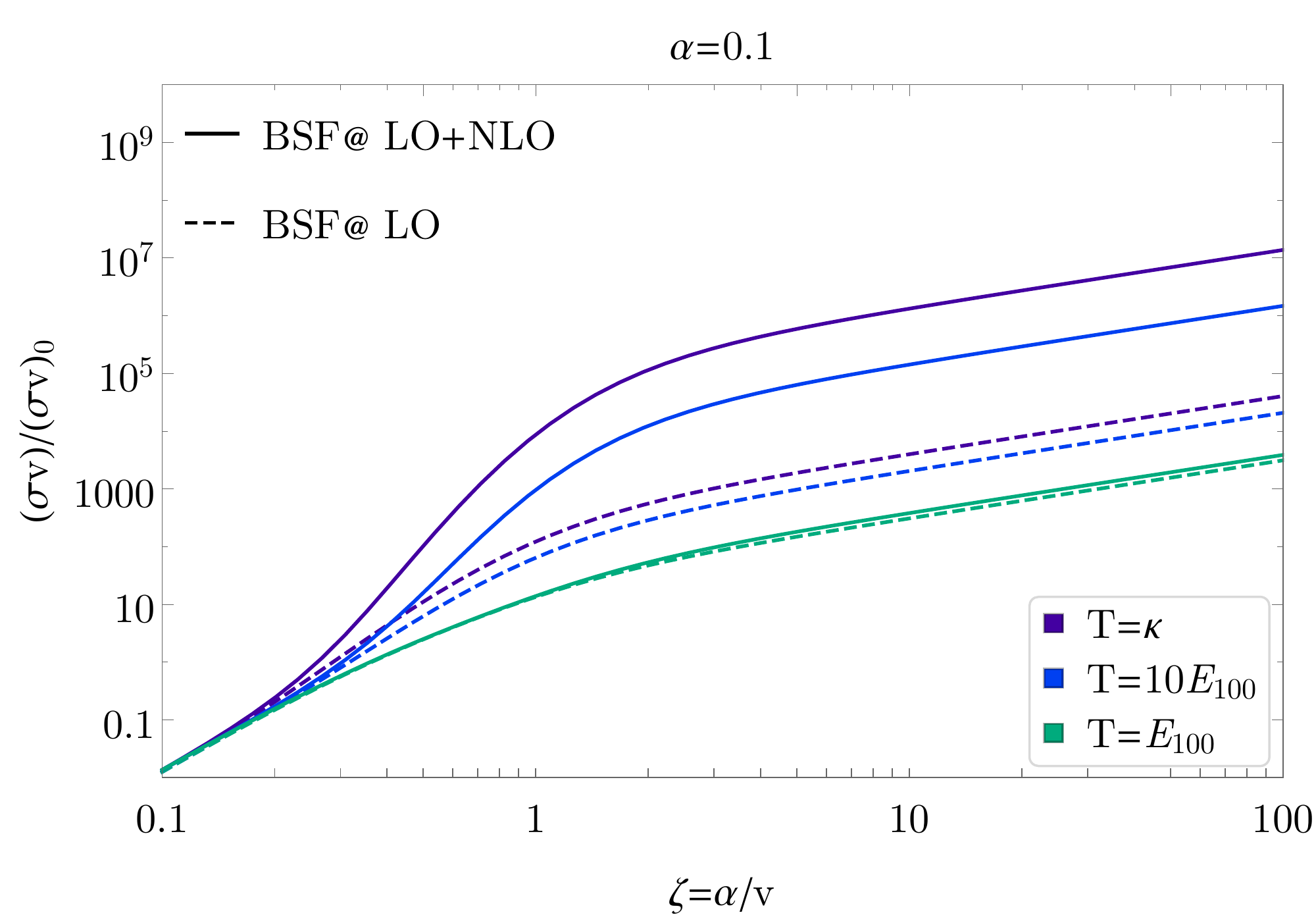}
\caption{Lower row shows the bound-state formation cross sections. Upper row compares the thermally averaged quantities. Validity of dipole approximation breaks down for $T \gtrsim \kappa= \mu \alpha$. The cross sections are normalized to  $(\sigma v)_0 \equiv  2 \pi \alpha^2/m_{\chi}^2 $. }
\label{fig:totalxsections}
\end{figure}

The leading order cross section for capture into the ground state $(\sigma^{\text{LO}}_{100} v_{\text{rel}})$ in Eq.~(\ref{eq:mBSFxsection}) and our result of the next-to-leading order cross section $(\sigma^{\text{NLO}}_{100} v_{\text{rel}})$ in Eq.~(\ref{eq:separation_vac_nonvac}) are compared to each other. The thermally averaged quantities are shown in the top row of fig.~\ref{fig:totalxsections}. Overall one can recognize that \emph{the NLO starts to dominate over the LO cross section for temperatures larger than the absolute value of the  ground state binding energy}. While the thermally averaged BSF cross section via the on-shell mediator emission increases for decreasing temperature, the NLO contribution becomes larger for higher temperature. 

The reason for this behavior can be understood from the lower panel of fig.~\ref{fig:totalxsections}, where  the non-averaged quantities as a function of inverse relative-velocity are shown. In the high temperature regime around the Bohr-momentum $\kappa$, the total cross section differs in shape and amplitude compared to the LO. The overall increase in the amplitude towards higher temperatures is mainly caused by the number density of relativistic bath particles, scaling as $T^3$. Additionally, the NLO cross section has a stronger dependence on the relative-velocity for $v_{\text{rel}} \gtrsim \alpha$, which turns into the milder scaling $v_{\text{rel}}^{-1}$ for $v_{\text{rel}} \ll \alpha$. The temperature and velocity enhancement leads to the fact that the overall BSF probability distribution in the thermal average is sharply peaked at the DM relative momentum $\sim \kappa$ and contributes less at the conventional one $\sim (m_{\chi}T)^{1/2}$. These features are less pronounced in the on-shell emission case, qualitatively explaining the differences of the thermally averaged quantities.

\subsection{When is a thermal field theory approach required?}
\label{sec:massive}

In the Boltzmann formalism, the bound-state formation cross section for bath-particle 
\begin{wrapfigure}{l}{0.5\textwidth}
\begin{center}
\includegraphics[scale=0.34]{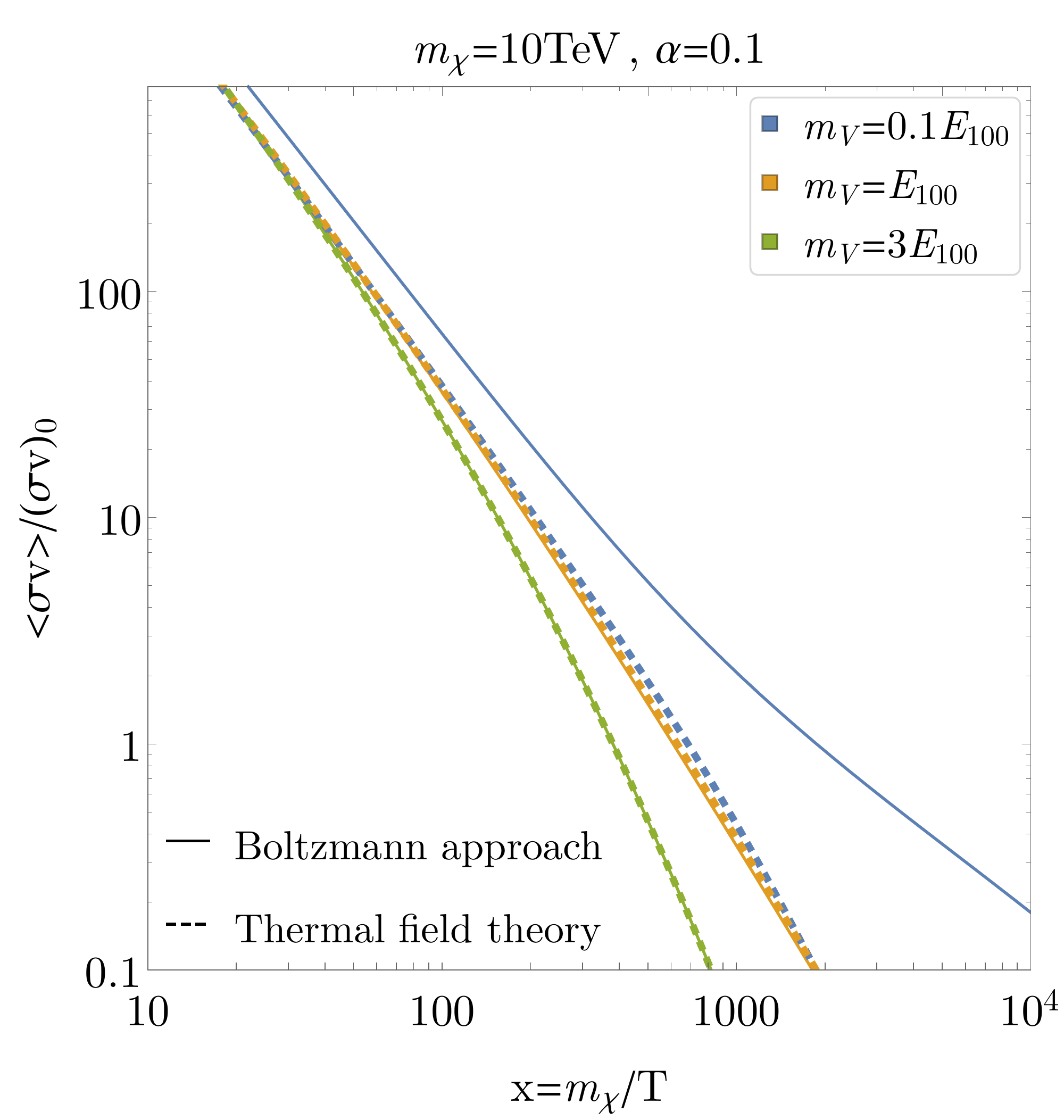}
\caption{Graphic compares the thermally averaged bound-state formation cross section for bath-particle scattering in the conventional Boltzmann and the thermal field theory approach.}
\label{fig:finitemass}
\end{center}
\end{wrapfigure}
scattering is always finite for massive mediators. However in the limit mass to zero, the cross section has a logarithmic divergence, which originates from the forward scattering divergence of the bath-particles, see also ref.~\cite{Binder:2019erp}. In this section, we would like to answer the following question for practical purposes: what is the critical mediator mass, above which the conventional Boltzmann formalism is a sufficient description and below which a more sophisticated thermal field theory analysis as presented in our work is required?

To estimate the critical mediator mass, only the dominant contributions of $R^{++}$ and $R^{--}$ are considered,  which contain the BSF via particle and anti-particle scattering, respectively. Additionally, these functions also contain interference terms originating from the double poles of the photon propagator. For a vector mediator with a mass $m_V$, one can simply replace the photon propagator by a massive one. This changes only the double pole into:
\begin{align}
G^{++}_{\epsilon}(|\mathbf{p}|,\tau,|\mathbf{k}|) = \frac{ F^{++}(|\mathbf{p}|,\tau,|\mathbf{k}|)}{\left([\Delta E+i \epsilon]^2 - \mathbf{p}^2 - m_V^2\right)^{2} \left(\Delta E + |\mathbf{k}| -  |\mathbf{p}+\mathbf{k}| + i \epsilon\right)},
\end{align}
where the single pole and $F^{++}$ remain unaffected. One can immediately recognize that the double pole at $|\mathbf{p}|=\tilde{z}_0= (\Delta E^2 - m_V^2)^{1/2}$ only exists for $\Delta E>m_V$. This is always true for mediator masses smaller than the absolute value of the binding energy. In general, for massive mediators one can replace $R^{++}$ by: 
\begin{align}
\lim_{\epsilon \searrow 0}R^{++}_{\epsilon} = \int_0^{\infty} \text{d} |\mathbf{k}| \int_{-1}^{1} \text{d} \tau \left[ \theta(\tilde{z}_0^2) \text{Res}(G_{0}^{++},\tilde{z}_0) + \text{Res}(G_{0}^{++},z_p) \right].\label{eq:massive}
\end{align}
As a reminder, the term $\text{Res}(G_{0}^{++},z_p)$ corresponds to BSF via bath-particle scattering and reproduces ref.~\cite{Binder:2019erp}. From this equation one can see that the double pole contribution, which causes the difference to the Boltzmann formalism, can be neglected for \emph{mediator masses much larger than the binding energy} due to $\theta(\tilde{z}_0^2)$. This statement is confirmed numerically as shown in fig.~\ref{fig:finitemass}, where we compute the BSF cross section based on the full Eq.~(\ref{eq:massive}) and without the double pole contribution. The Boltzmann formalism overestimates the effect of the bath-particle scattering for $m_V \ll  E_{100}$. For $m_V \gg  E_{100}$, the Boltzmann formalism is reliable.

\section{Impact on thermal relic abundance}
\label{sec:comparison}
The results of the previous sections allow us to explore the impact of higher-order BSF processes on the evolution of the thermal relic abundance for the following simplified model with $N$ bath particle species:
\begin{align}
\mathcal{L} \supset -g \bar{\chi} \gamma^{\mu} \chi A_{\mu} - \sum \displaylimits_{i=1}^{N} g \bar{\psi}_i \gamma^{\mu} \psi_i A_{\mu}. \label{eq:darkQED_lightFermions}
\end{align}
An s-wave $\chi$-pair can annihilate into $2 \gamma$ and N effectively massless $\psi$ pairs.
The total s-wave Sommerfeld enhanced annihilation cross section, averaged over the initial spin and summed over final degrees of freedom, can be written as
\begin{align}
\sigma^{\text{an}} v_{\text{rel}}  &= \left(1+N\right)\frac{\pi \alpha^2}{m_{\chi}^2}  |\psi_{\mathbf{k},l=0}(r=0)|^2,
\end{align}
where the first two factors account for the tree-level part and $|\psi_{\mathbf{k},l=0}(0)|^2=2\pi \zeta(1-e^{-2 \pi \zeta})^{-1}$ is the Sommerfeld enhancement factor~\cite{doi:10.1002/andp.19314030302, Sakharov:1948yq}. For the lowest s-wave $\chi$ bound states,
we consider the spin singlet (S) decay into $2 \gamma$~\cite{pirenne1946proper,doi:10.1111/j.1749-6632.1946.tb31764.x}, and the spin triplet decay into $3\gamma$~\cite{1949PhRv...75.1696O} and N $\psi$-pairs~\cite{Bilenky:1969zd}. The decay widths can be written as
\begin{align}
\Gamma_{S}^{\text{dec}} &= 4 \times \frac{\pi \alpha^2}{m_{\chi}^2} |\psi_{100}(r=0)|^2, \\
\Gamma_{T}^{\text{dec}} &= \left(c_{3 \gamma} + \frac{N}{3} \right)\Gamma_{S}^{\text{dec}},
\end{align}
where $|\psi_{100}(0)|^2 = (\mu \alpha)^3/\pi$, and $c_{3 \gamma} = 4(\pi^2 - 9)\alpha/(9\pi)$~\cite{1949PhRv...75.1696O}. We compute the BSF cross section in Eq.~(\ref{eq:generalcrosssection}) for the capture into the ground state up to NLO in the dark photon spectral function:
\begin{align}
\langle  \sigma^{\text{bsf}}_{100} v_{\text{rel}}\rangle  \simeq \langle  \sigma^{\text{LO}}_{100} v_{\text{rel}}\rangle + N \langle \sigma^{\text{NLO}}_{100} v_{\text{rel}}\rangle .
\end{align}
For the LO cross section we can directly use the expression in Eq.~(\ref{eq:mBSFxsection}), while for the NLO contribution we take $N$ times our results in Eq.~(\ref{eq:separation_vac_nonvac}). The number density evolution of the lowest bound states can be included into the Boltzmann Eq.~(\ref{eq:gen_Boltz}) for the scattering states approximately as~\cite{vonHarling:2014kha,Ellis:2015vna}:\footnote{For the numerical solution of this equaion, we use a private mathematica version of the DarkSUSY~\cite{Bringmann:2018lay} relic density routine. For the case $N=0$, we reproduced the results of ref.~\cite{vonHarling:2014kha} as a check.}
\begin{align}
\dot{n}_{\chi} + 3 Hn_{\chi} =& 
- \left[\langle \sigma^{\text{an}} v_{\text{rel}} \rangle+ W(T) \right]
\left[n_{\chi}n_{\bar{\chi}}- n_{\chi}^{\text{eq}}n_{\bar{\chi}}^{\text{eq}} \right], \label{eq:numberdensity}\\
W(T) \equiv &  
\frac{\langle  \sigma^{\text{bsf}}_{100} v_{\text{rel}}\rangle }{4} \left[
\frac{\Gamma_{S}^{\text{dec}}}
{\Gamma_{S}^{\text{dec}} +  \Gamma_{S}^{ \text{dis}}} 
+
3\times \frac{ \Gamma_{T}^{\text{dec}}}
{\Gamma_{T}^{\text{dec}} +  \Gamma_{T}^{ \text{dis} }} 
\right].
\end{align}
The effective cross section $W(T)$ comprises all information of the bound states.
It consists out of a sum over the individual singlet and triplet BSF cross sections (here spin-independent), weighted by branching ratios containing the individual decay and dissociation rates. The latter corresponds to the inverse process of the bound-state formation and is therefore related via detailed balance as
\begin{align}
\Gamma_{i}^{\text{dis}} =&  \frac{\langle  \sigma^{\text{bsf}}_{100} v_{\text{rel}}\rangle }{4} \frac{ n_{\chi}^{\text{eq}}n_{\bar{\chi}}^{\text{eq}}}{n_{100}^{\text{eq}}}\;,
\end{align}
where the equilibrium number density of the bound state contains one spin d.o.f.

\begin{figure}
\centering
\includegraphics[scale=0.36]{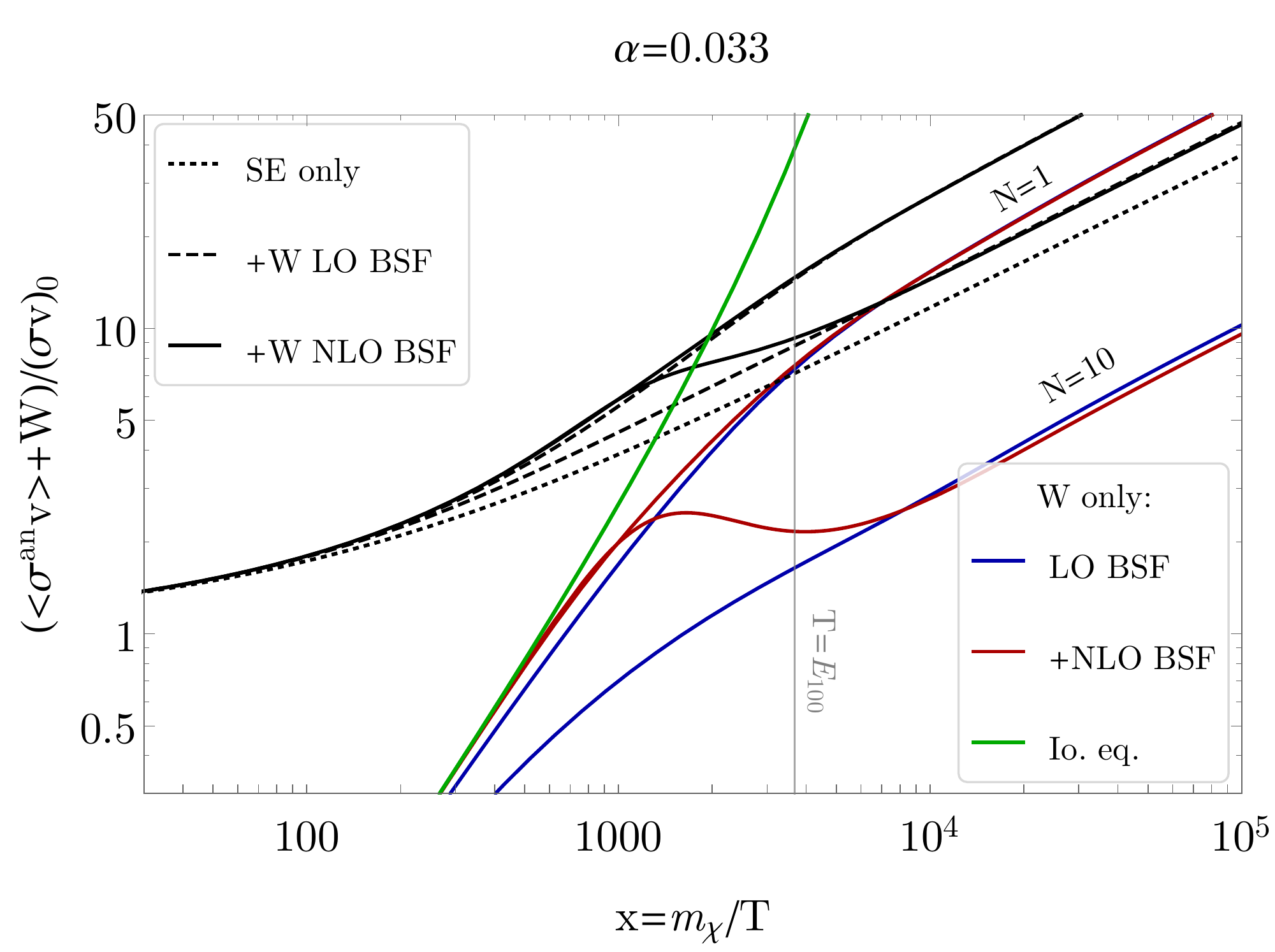}
\includegraphics[scale=0.36]{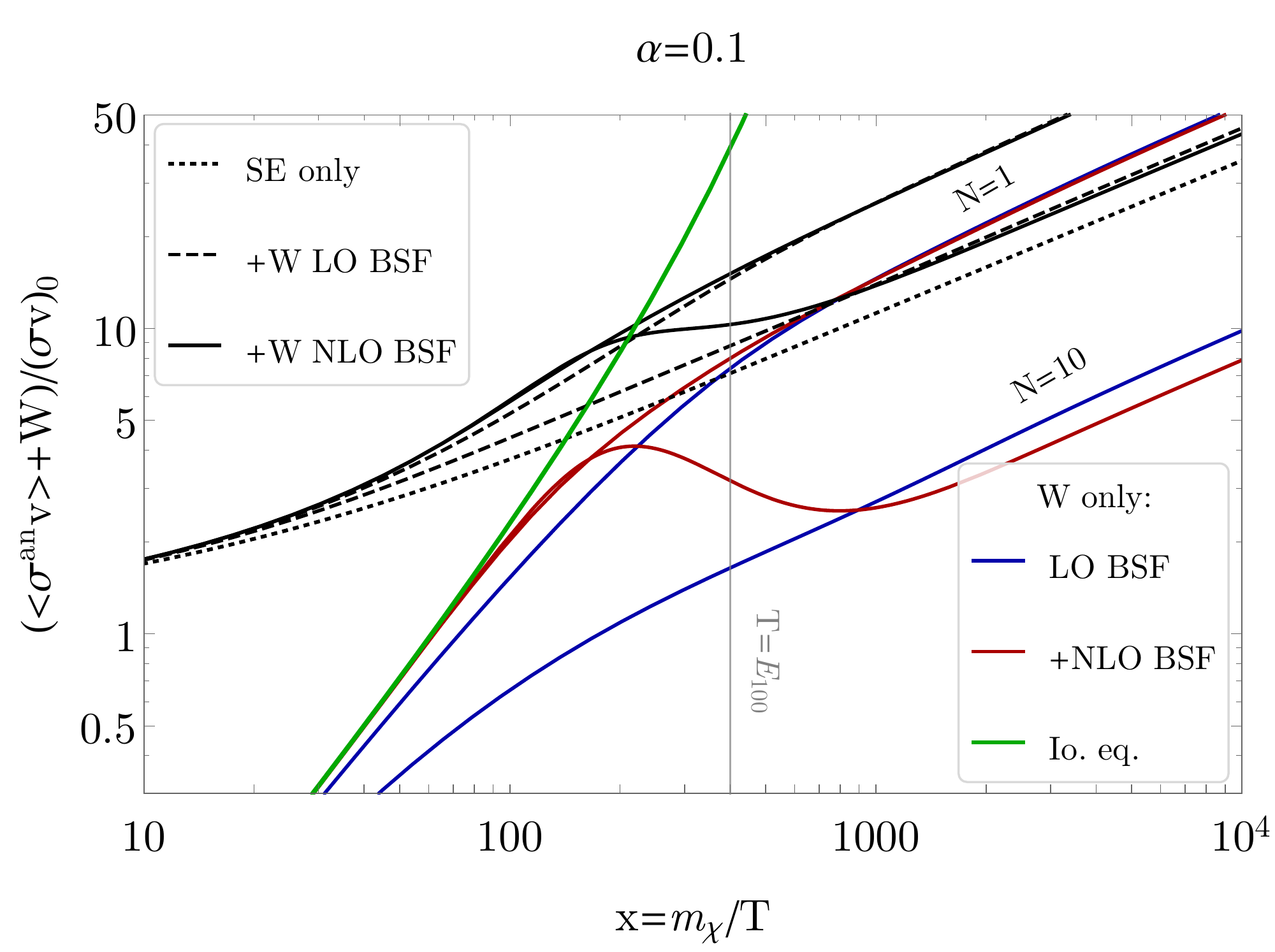}
\caption{Graphic shows the increasing importance of NLO BSF processes for larger $N$. Upper (lower) solid and dashed black line, as well as upper (lower) red and blue line correspond to the $N=1$ ($N=10$) case. All quantities are normalized to $(\sigma v)_0 \equiv  \left(1+N\right)\pi \alpha^2/m_{\chi}^2 $, such that the curve for SE annihilation and for $W$ in ionization equilibrium are independent of $N$.}
\label{fig:totalcollisionterm}
\end{figure}

The effective cross section $W$ features two asymptotic regimes:
\begin{equation}
W(T)\simeq \begin{cases*}
       (\Gamma_{S}^{\text{dec}}+ 3 \Gamma_{T}^{\text{dec}}) n_{100}^{\text{eq}}/(n_{\chi}^{\text{eq}}n_{\bar{\chi}}^{\text{eq}}) & \text{for} $\Gamma_{i}^{\text{dec}}\ll \Gamma_{i}^{\text{dis}}$ (ionization equilibrium), \\
      \langle\sigma^{\text{bsf}}_{100} v_{\text{rel}}\rangle & \text{for} $\Gamma_{i}^{\text{dec}}\gg \Gamma_{i}^{\text{dis}}$ (out-off io. eq.).
   \end{cases*}
   \label{eq:W-limit}
\end{equation}
In the first asymptotic regime, the bound and scattering states are in \emph{ionization equilibrium} (io.~eq.). Here, $W$ is i) \emph{independent} of the BSF cross section and dissociation rate~\cite{Binder:2018znk}, and ii) \emph{maximum} for a given bound-state decay rate and temperature. In fig.~\ref{fig:totalcollisionterm}, the maximum value of $W$ is indicated by the green line for electroweak (left panel) and strong couplings (right panel). In the second asymptotic regime at later times, the number density depletion depends only on the total BSF cross section, since the bound states immediately decay without being dissociated back into the scattering states (BSF without ``back reaction''). The transition between these two regimes occurs when the dissociation rate is comparable to the decay rates.

We estimate the maximum relative size of $W$ with respect to the SE annihilation cross section \footnote{In the instantaneous approximation of the SE, one can estimate $\langle\sigma^{\text{an}} v_{\text{rel}}\rangle \approx (1+N)(\sigma v_{\text{rel}})_0 (x/x_0)^{1/2}$ for $x \gg x_0$ and $x_0=3/(4 \pi^2 \alpha^2)$, see, e.g., appendix in ref.~\cite{Binder:2017lkj} for details.}, assuming io.~eq.:
\begin{align}
\frac{W(T)}{\langle\sigma^{\text{an}} v_{\text{rel}}\rangle} \approx 2 \times \frac{|E_{100}|}{T} e^{|E_{100}|/T}. \label{eq:Wanratio}
\end{align}
In this model, the ratio is independent of the number $N$ of bath-particle species. In io.~eq., $W$ first grows rapidly for decreasing temperature with a power law $(m_{\chi}/T)^{3/2}$, becomes comparable to the SE for temperature around twice the binding energy, and has an exponentially growing factor for lower temperature.

The relevance of NLO BSF contributions depends, for instance, on for how long the LO BSF can keep the system in ionization equilibrium. In fig.~\ref{fig:totalcollisionterm}, we show that for only one bath-particle species in the plasma, the on-shell emission of a vector mediator is still effective enough to keep $W$ close to its maximum value for sufficiently long time. In this case, the effect of the NLO contributions is only marginally relevant. For larger $N$, however, the relevance of the NLO contributions is more important. This is because the SE annihilation, NLO BSF cross section, and triplet decay rate are proportional to $N$, while the LO BSF is independent. BSF via bath-particle scattering as the dominant NLO process, can keep the system in io.~eq. until temperatures close to the binding energy even for large $N$.

\begin{figure}
\centering
\includegraphics[scale=0.29]{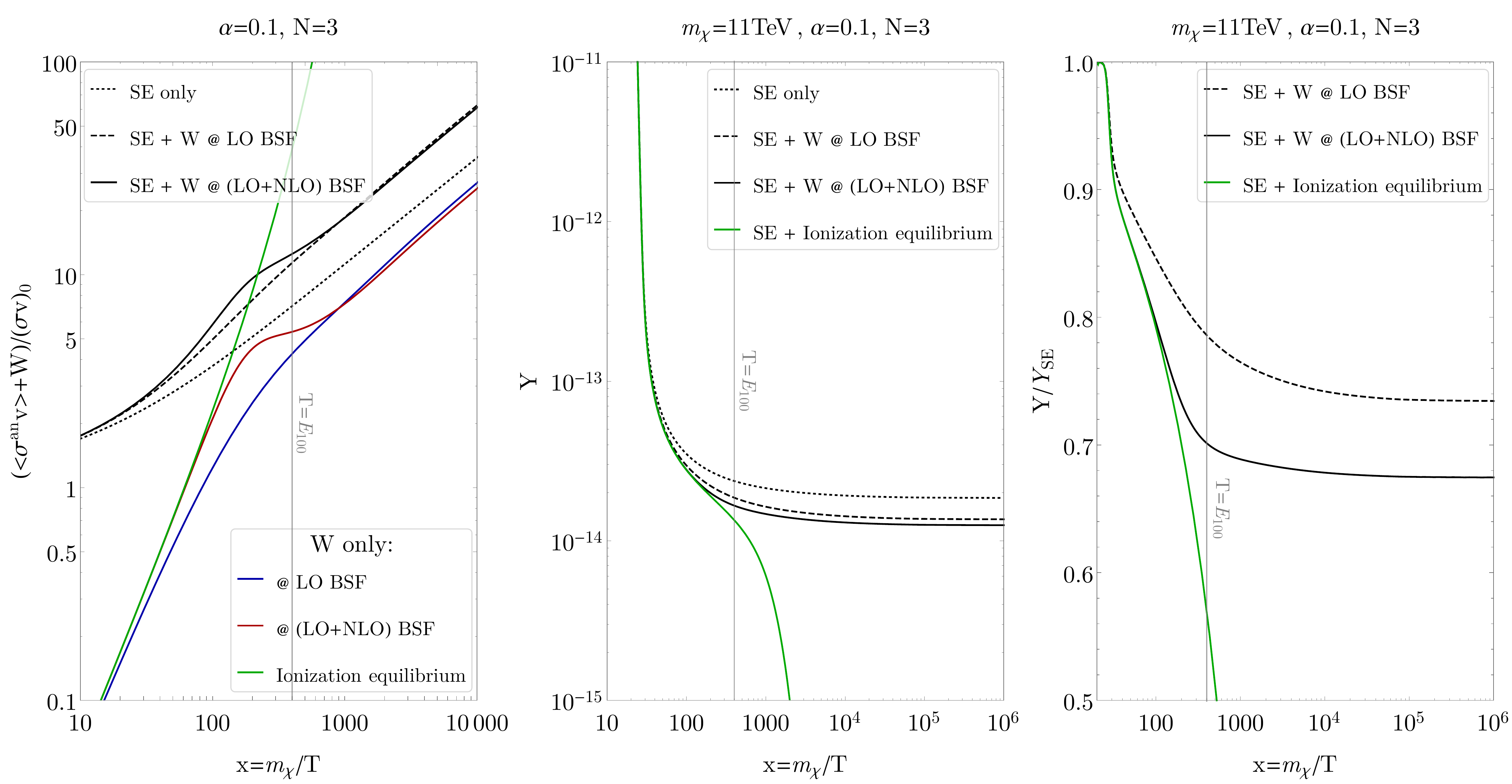}
\caption{Graphics show the impact of bound-state formation at NLO on the evolution of the thermal relic abundance. Here, $Y\equiv n_{\chi}/s$, where $s$ is the SM entropy density, and $Y_{\text{SE}}$ assumes Sommerfeld enhanced annihilation only.}
\label{fig:relicabundance}
\end{figure}

Moreover, the relevance of NLO BSF contributions depends on the relative size of $W$ compared to the SE annihilation around the decoupling time from ionization equilibrium. Close to the binding energy, the depletion of the DM number density is exponentially sensitive to the ionization decoupling temperature according to Eq.~(\ref{eq:Wanratio}). While in the conventional Boltzmann framework one can artificially push the decoupling temperature below the binding energy by lowering the vector mediator mass, our thermal field theory result suggests even for a massless vector mediator a decoupling before but close to the strong exponentially enhanced regime. Based on this result, we may expect a significant impact on the predicted relic abundance in the model under consideration mainly for strong couplings $\alpha \sim 0.1$.

\begin{figure}
\centering
\includegraphics[scale=0.47]{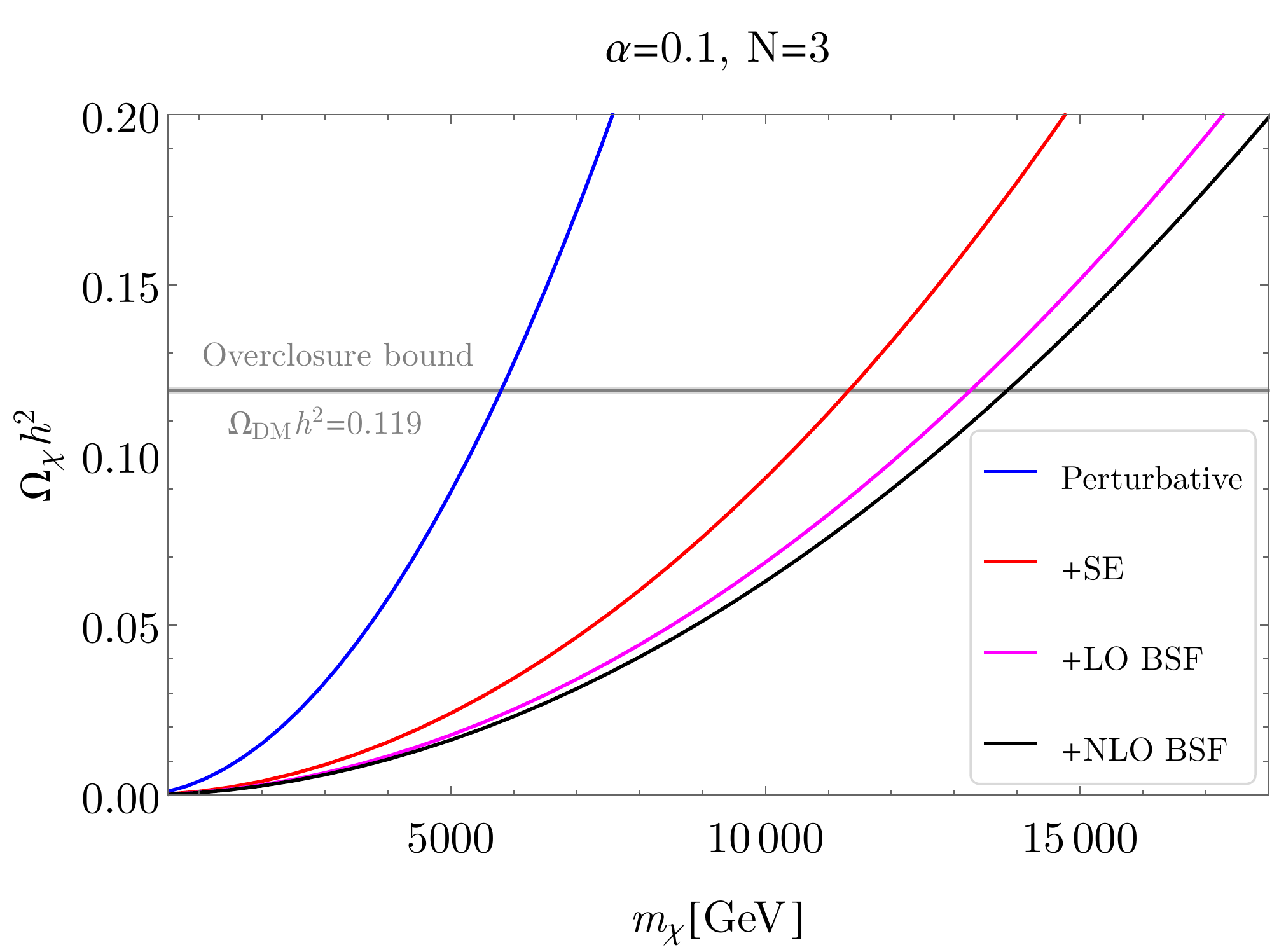}
\caption{Under the assumption of a strong coupling $\alpha=0.1$ and one quark in the plasma, we show the predicted relic abundance when considering a tree-level cross section only (blue), including the Sommerfeld effect (red), considering BSF via LO on-shell emission (pink), as well as BSF at NLO (black).}
\label{fig:overclosure}
\end{figure}

To demonstrate this more clearly, we choose a strong coupling value and more conservatively $N=3$ (``one quark'') in fig.~\ref{fig:relicabundance}. Here, the effect of NLO corrections can significantly change the relic abundance due to a delayed decoupling from ionization equilibrium. The resulting corrections to the upper limit of the DM mass, above which the Universe would contain too much DM (overclosure bound), are shown in fig.~\ref{fig:overclosure}. For choices of larger $N$, the differences between LO and NLO would further increase.

\section{Discussion}
\label{sec:dis}

From the results of previous sections, we learned that DM bound-state formation inside a relativistic thermal bath can be dominated by higher order processes for a certain temperature range for massless mediators. It was demonstrated that next-to-leading order processes contained in the mediator spectral function have the potential to change the thermal relic abundance significantly. Although we analyzed a simplified model, this message should be recognized in the broader context of coannihilation scenarios, where the number of SM bath particles can be much larger than considered here in this work. 

The formalism of this work also allows to study the impact of the ambient primordial plasma on excited DM state transitions. One can simply change the initial scattering state into a bound state, which transforms the BSF cross section Eq.~(\ref{eq:generalcrosssection}) into a level-transition rate. The factorization between the spectral function and the level-transition matrix element holds in this case as well. Contributions from higher DM states were however dropped in the previous section and remain a major uncertainty in the prediction of the final relic abundance.

The bound-state formation and the reverse process have also been investigated in the context of heavy  annihilating quarkonia inside the quark-gluon plasma produced, e.g., at the Large-Hadron-Collider, which has many similarities compared with heavy annihilating dark matter in the primordial plasma. For fairness, we  compare our approach to the existing methods in the quarkonia literature in the following. A similar formalism is developed in ref.~\cite{Yao:2018nmy}, where the LO cross section for quarkonium formation and dissociation is derived from the Lindblad equation. In ref.~\cite{Yao:2018sgn}, 
a particular part of cancellation of collinear divergences is suggested while an ad hoc insertion of a finite temperature self-energy into vacuum amplitudes was performed. Our work can be regarded as a first complete analysis at NLO in the mediator spectral function. Starting from first principles, we have derived Eq.~\eqref{eq:generalcrosssection} and demonstrated that the mediator spectral function evaluated at NLO automatically leads to a collinear safe cross section and includes all the possible processes simultaneously. For example, we have shown that even the finite part of the interference terms is important for predicting the relic abundance precisely (see Fig.~\ref{fig:rfunction}), which cannot be addressed without the full NLO treatment.

The limitation of the formalism presented in this work is set by the assumptions of potential non-relativistic effective field theory, including the leading order dipole approximation. For temperatures larger than the Bohr-momentum $\kappa$, the plasma can entirely probe the typical size of the bound state, and the validity of our generalized BSF cross section Eq.~(\ref{eq:generalcrosssection}) breaks down.
At the critical value $T \sim \kappa$, the bound and scattering states become non-separable because of the rapid transition between them, which calls for another formalism capable of coping with this situation.
At the same time, it was shown that BSF and the back reaction is such efficient, that the system is robustly in ionization equilibrium. 
We emphasize that the collision term is independent of the BSF cross section in such a state (see Eq.~\eqref{eq:W-limit}). 

\begin{figure}
\centering
\includegraphics[scale=0.8]{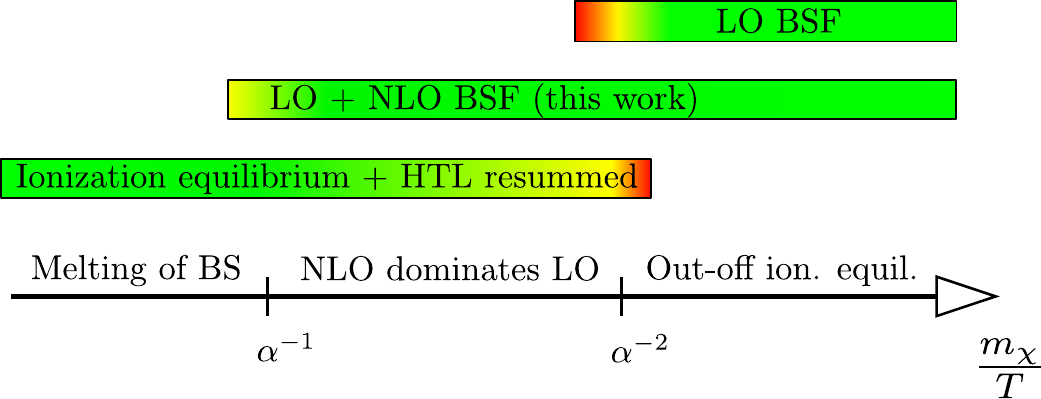}
\caption{Graphic illustrates goodness of different descriptions, based on the assumption that the mediator is coupled to the primordial plasma background. LO and LO+NLO stand for the bound-state formation description via the emission of an on-shell mediator and higher order couplings of the mediator to the plasma, respectively. The bottom bar line represents HTL resummed corrections to the SE annihilation and bound-state decay rate under the assumption that ionization equilibrium is maintained.}
\label{fig:overview}
\end{figure}

By using another formalism valid at high temperature but limited to ionization equilibrium, environmental corrections to the upper formula of \eqref{eq:W-limit} have been studied in the literature~\cite{Biondini:2017ufr,Biondini:2018xor,Biondini:2018pwp,Biondini:2018ovz,Kim:2019qix,Biondini:2019int}. The method is based on an effective in-medium potential, which shows next to the expected Debye screening mass, a temperature dependent energy shift (Salpeter correction), as well as an imaginary thermal width~\cite{Laine:2006ns,Brambilla:2008cx}. A conservative estimate shows that the thermal width in the effective in-medium potential can lead to an entire \emph{melting} of all bound states for $T \gtrsim \kappa$~\cite{Kim:2016kxt,Kim:2016zyy}. Such melting phenomena have been experimentally observed, e.g., in the decay spectra of bottomonium systems in a quark-gluon plasma, see fig.1 in ref.~\cite{Sirunyan:2018nsz}. In ref.~\cite{Binder:2018znk}, it has been  emphasized that the formalism for computing the relic abundance in the effective in-medium potential description is limited to the assumption of ionization equilibrium (see also ref.~\cite{Biondini:2019zdo}), illustrated by the bottom bar line in fig.~\ref{fig:overview}.

One of the main formal achievements of this work was to develop a complementary method, which overlaps with the validity region of the previous effective in-medium potential approach. Combined, we can now describe the evolution of the DM system for a broader range of temperatures, ranging from the melting of bound states in the high temperature regime down to arbitrarily below the decoupling from ionization equilibrium. In particular, we have now addressed the temperature regime below the Bohr-momentum, where the thermal width can be treated as a perturbation and bound states can be characterized by their usual quantum numbers. Note again that the expansion given in Eq.~\eqref{eq:twobodyexp} implicitly assumes a clear separation between the scattering and bound states, which is a crucial assumption when the continuum starts to overlap with the discrete energy spectrum in the high temperature regime. For lower temperatures around the binding energy, these corrections are expected to be negligible compared to our improved description of the decoupling from ionization equilibrium. In our model, we found that the depletion of the dark matter density is \emph{exponentially sensitive} to the precise decoupling temperature from ionization equilibrium. The presented formalism in this work now allows to accurately resolve the number density evolution especially during this regime.

\section{Summary and conclusion}
\label{sec:con}

In the conventional Boltzmann formalism, the amplitude for higher-order bound-state formation processes can become collinear divergent in the case of massless mediators. A large number of possible bound-state formation channels through SM particle scattering, via the virtual exchange of, e.g., photons or gluons in coannihilation scenarios, can not be investigated in this context. Based on non-equilibrium quantum field theory techniques, we derived a more general cross section, Eq.~(\ref{eq:generalcrosssection}), which addressed this issue without the need of an ad hoc screening mass regulator. We presented for the first time a full computation and analysis of the thermal one-loop correction for the dominant capture into the ground state in the case of ultra-relativistic fermions in the primordial plasma environment, resembling the interactions with light SM leptons and quarks. For temperatures larger than the absolute value of the binding energy, we found that bound-state formation via bath-particle scattering dominates also for massless vector mediators over the so far considered on-shell mediator emission. Our results are complementary to the one in ref.~\cite{Binder:2019erp}, where massive mediators where investigated.

The key quantity in the generalized bound-state formation cross section is the spectral two-point correlation function of the interacting mediator. It was demonstrated that a perturbative expansion in the coupling parameter of this spectral function successively generates on-shell and virtual mediator contributions in a proper thermal field theoretical framework. The known result for the capture into the ground state via the emission of an on-shell mediator was reproduced at the lowest order in the perturbative expansion. The higher-order BSF processes via bath-particle scattering and off-shell mediator decay were identified at the first interaction level. These processes are collinear divergent for massless mediators in the conventional Boltzmann approach. It was shown that other terms automatically occur in the spectral function in addition, canceling the collinear divergences and resulting in a finite collision term. A rather general analytic proof for the collinear finiteness, applying in particular to our full first interaction term, was presented in section~\ref{sec:cancellation}. 

Based on our extended analysis in section~\ref{sec:massive}, we conclude that a thermal field theory approach is required for models where the mediator mass is smaller than the absolute value of the binding energy. This implies that our approach can become important also for massive SM gauge bosons or the Higgs field during the electroweak cross over, featuring a large number of possible BSF channels via SM particle scattering as well.

In the case of our simplified model, we demonstrated in section~\ref{sec:comparison} that the impact of the new higher-order bound-state formation effects on the thermal relic abundance can be especially significant if the on-shell emission is not enough to keep the system in ionization equilibrium until temperatures close to the binding energy. Regarding more realistic coannihilation scenarios, the size of the corrections to the upper bound on the DM mass from higher-order bound-state formation processes still remains an open question from the perspective of this work, since the details of the underlying model might play a crucial role. 
This is mainly due to the fact that the impact on the relic abundance is exponentially sensitive to the precise decoupling time from ionization equilibrium if it is maintained until temperatures around the binding energy. Model-dependent order one variations in the ionization equilibrium decoupling temperature are crucial for the impact of the higher-order effects. With an accurate formalism by hand, together with the insights from simplified DM models, makes it nevertheless worthwhile to analyze more realistic scenarios in future work.

\acknowledgments

We are thankful to Kalliopi Petraki for her initial contribution at the early stage of this work.
T.B. was supported by World Premier International Research Center Initiative (WPI), MEXT, Japan.
J.H. is supported by the DFG Emmy Noether Grant No. HA 8555/1-1.
We also would like to thank Laura Covi, Peter Cox, Bj\"orn Garbrecht, Paolo Gondolo, Felix Kahlhoefer, Karol Kova\v{r}\'{i}k, Feng Luo, Shigeki Matsumoto, Satoshi Shirai, and Carlos Tamarit for discussion on this subject.


\appendix

\section{Computation of double commutator}
\label{app:doublecommutator}

In this appendix, we give a detailed derivation of Eq.~\eqref{eq:BSFcollisionterm} starting from Eq.~\eqref{eq:vanNeumann}.
Throughout this paper, we are interested in the BSF and dissociation rate.
For this purpose, one may focus on a particular part of the whole interaction Hamiltonian,
which is nothing but the dipole interactions given in Eq.~\eqref{eq:dipolop}.
Inserting the mode expansion given in Eq.~\eqref{eq:twobodyexp}, we obtain the following form for the operator responsible for the transition between scattering and bound states:
\begin{align}
	\hat H_{\text{dip}} \supset 
	- \int\limits_{\mathbf{K},\mathbf{k},P} \sum_{\mathcal{B},\text{Spin}} e^{i(\Delta E-P_{0})t} \; E^{i} (P)  L_{\mathbf{k},\mathcal{B}}^{i\, \star} \;\hat a^{s \dagger}_{\mathbf{K}/2+\mathbf{k}} \hat  b^{r \dagger}_{\mathbf{K}/2-\mathbf{k}} \hat c^{sr}_{\mathcal{B},\mathbf{K}} + \text{H.c.},
	\label{eq:dipole-app}
\end{align}
where the overlap integral is given by
\begin{align}
	L_{\mathbf{k},\mathcal{B}}^{i} = g \int \dd^{3}r \; 
	\psi^{\star}_{\mathcal{B}}(\mathbf{r}) r^{i} \psi_{\mathbf{k}}(\mathbf{r}).
\end{align}
Here we use a shorthand notation for integrals:
$\int_\mathbf{k} \equiv \int \frac{\dd^3 k}{(2\pi)^3}$ and 
$\int_P \equiv \int \frac{\dd^4 P}{( 2 \pi )^4}$.
The positive quantity $\Delta E \equiv {\mathcal E}_{\mathbf{k}} - {\mathcal E}_{\mathcal{B}}$ is the total energy emitted in the inelastic conversion, i.e.\ relative kinetic energy plus the absolute value of the negative binding energy.

All one needs to do is to evaluate the double commutator in the right-hand side of Eq.~\eqref{eq:vanNeumann} by using Eq.~\eqref{eq:dipole-app} in order to obtain the collision term for the BSF and dissociation rates.
Below we summarize commutators and expectation values relevant for our computations:
\begin{align}
	\left[ \hat n_{\mathbf{k}_{\chi}}, \hat a_{\mathbf{k}'_{\chi}}^{s} \right] = - (2\pi)^3\delta^{3} (\mathbf{k}_{\chi} - \mathbf{k}'_{\chi}) \hat a_{\mathbf{k}'_{\chi}}^{s}, \qquad
	\left[ \hat n_{\mathbf{k}_{\chi}}, \hat a_{\mathbf{k}'_{\chi}}^{s \dag} \right] = (2\pi)^3 \delta^{3} (\mathbf{k}_{\chi} - \mathbf{k}'_{\chi}) \hat a_{\mathbf{k}'_{\chi}}^{s \dag},
\end{align}
for commutators, and
\begin{align}
	& \left< a_{\mathbf{k}_{\chi}}^{s \dag} a_{\mathbf{k}'_{\chi}}^{s'} \right> = f_{\chi} (\mathbf{k}_{\chi}) (2\pi)^3 \delta^{3} (\mathbf{k}_{\chi} - \mathbf{k}'_{\chi}) \delta_{s s'}, \quad
	\left< b_{\mathbf{k}_{\bar\chi}}^{s \dag} b_{\mathbf{k}'_{\bar\chi}}^{s'} \right> = f_{\bar\chi} (\mathbf{k}_{\bar\chi}) (2\pi)^3 \delta^{3} (\mathbf{k}_{\bar\chi} - \mathbf{k}'_{\bar\chi}) \delta_{s s'}, \nonumber\\
	&\left< c_{\mathcal{B},\mathbf{K}}^{sr \dag} c_{\mathcal{B}',\mathbf{K}'}^{s' r'} \right> = f_{\mathcal{B}} (\mathbf{K}) (2\pi)^3 \delta^{3} ( \mathbf{K} - \mathbf{K}' ) \delta_{\mathcal{B} \mathcal{B}'} \delta_{s s'} \delta_{r r'},
\end{align}
for expectation values of creation and annihilation operators.
Here note that these expectation values are taken by the initial factorized density matrix:
$\langle \hat\bullet \rangle = \Tr [\hat\bullet \hat \rho (t = 0)]$.
We also need two-point functions of the electric field defined by
\begin{align}
	E_{ij}^{-+}(P) &\equiv \int \text{d}^4(x-y) e^{- iP \cdot (x-y)} \left< E^i(x) E^{j}(y) \right>, \quad \nonumber \\
	E_{ij}^{+-}(P) &\equiv \int \text{d}^4(x-y) e^{- iP \cdot (x-y)} \left< E^{j}(y) E^i(x)  \right>.
\label{eq:ECorrelator}
\end{align}
For later convenience, we rewrite this electric correlator in terms of its gauge field as:
\begin{align}
	E^{-+/-+}_{ij}(P) = (i p^0 g^{i\mu} - i p^i g^{0 \mu}) ( - i p^0 g^{j\mu} + i p^j g^{0 \nu})  D^{-+/+-}_{\mu \nu}(P),\label{eq:Epotrel}
\end{align}
where in coordinate space $D^{-+}_{\mu\nu}(x - y) \equiv \langle A_\mu (x) A_\nu (y) \rangle$ and $D^{+-}_{\mu\nu}(x - y) \equiv \langle  A_\nu (y) A_\mu (x) \rangle$. 
Since the expectation value is taken by the factorized initial density matrix and the environment is assumed to be in thermal equilibrium, these correlators coincide with the thermal correlator, fulfilling the KMS relation:
\begin{align}
D^{-+}_{\mu\nu}(P) = [1+f_{\gamma}^\text{eq}(P^0) ] D^{\rho}_{\mu\nu}(P), \quad D^{+-}_{\mu\nu}(P) &= f_{\gamma}^\text{eq}(P^0) D^{\rho}_{\mu\nu}(P),
\end{align}
where the photon spectral function is given by $D^{\rho}_{\mu\nu}(x - y) \equiv \langle [A_\mu (x), A_\nu (y)] \rangle$.
$f_{\gamma}^\text{eq}$ represents the Bose-Einstein distribution for the gauge boson.

Using these equations, one finally obtain the collision term associated with the BSF and dissociation:
\begin{align}
	\mathcal{C}_{\text{BSF} + \text{diss}} = 
-g_{\chi}g_{\bar{\chi}}\sum_{\mathcal{B}}  &\int \frac{\text{d}^3 k_{\chi}}{(2 \pi)^3}\frac{\text{d}^3 k_{\bar{\chi}}}{(2 \pi)^3}\frac{\text{d}^3p}{(2 \pi)^3}  D^{\rho}_{\mu \nu}(\Delta E,\mathbf{p})\sum_{\text{Spin}} \mathcal{T}^{\mu}_{\mathbf{k},\mathcal{B}} (\Delta E, \mathbf{p}) \mathcal{T}^{\nu \star}_{\mathbf{k},\mathcal{B}}(\Delta E, \mathbf{p}) \nonumber \\
&\times\bigg\{ f_\chi(\mathbf k_{\chi}) f_{\bar{\chi}}(\mathbf{k}_{\bar{\chi}}) [1+f_{\gamma}^\text{eq}(\Delta E) ]   - f_{\mathcal{B}}(\mathbf{K} - \mathbf{p}) f_{\gamma}^\text{eq}(\Delta E) \bigg\},
\label{eq:CollisionTermApp}
\end{align}
with $\Delta E =  \mathbf{k}^2/(2\mu)+|\mathcal{E}_{\mathcal{B}}|$, $\mathbf{k} = (\mathbf{k}_{\chi} - \mathbf{k}_{\bar{\chi}})/2$, and $\mathbf{K} = \mathbf{k}_\chi + \mathbf{k}_{\bar\chi}$.
Here we have dropped the Bose enhancement and Fermi suppression factors for the scattering and bound states because they are dilute.
To obtain this compact form, we have used the following relation between the overlap integral and tensors defined in Eqs.~\eqref{eq:tmupnr2} and \eqref{eq:tmupnr}:
\begin{align}
\frac{1}{g_{\chi} g_{\bar{\chi}}} E_{ij}^{\rho}(P) \sum_{\text{Spin}} L_{\mathbf{k},\mathcal{B}}^{i} \delta^{ss^{\prime}}\delta^{rr^{\prime}}  L_{\mathbf{k},\mathcal{B}}^{j \star} \delta^{ss^{\prime}}\delta^{rr^{\prime}}= D_{\mu \nu}^{\rho}(P) \sum_{\text{Spin}}  \mathcal{T}_{\mathbf{k},\mathcal{B}}^{\mu}(P) \mathcal{T}_{\mathbf{k},\mathcal{B}}^{\nu \star}(P),\label{eq:relEPOT}
\end{align}
where $g_\chi$ and $g_{\bar \chi}$ represent the spin degrees of freedom for particle and anti-particle respectively.
The collision term can be expressed as the following simple form
\begin{align}
	\mathcal{C}_{\text{BSF} + \text{diss}} = -\sum_{\mathcal{B}} \langle \sigma_{\mathcal{B}}^{\text{bsf}} v_{\text{rel}} \rangle \left[n_{\chi} n_{\bar{\chi}} - n_{\mathcal{B}} \frac{n_{\chi}^{\text{eq}} n_{\bar{\chi}}^{\text{eq}}}{n_{\mathcal{B}}^{\text{eq}}} \right],
	\label{eq:gen_Boltz-app}
\end{align}
where
\begin{align}
	\sigma_{\mathcal{B}}^{\text{bsf}} v_{\text{rel}} &\equiv \int \frac{\text{d}^3 p}{(2\pi)^3} \left[ 1 + f_\gamma^\text{eq} (\Delta E) \right] D^{\rho}_{\mu \nu}(\Delta E,\mathbf{p}) \sum_{\text{Spin}} \mathcal{T}^{\mu}_{\mathbf{k},\mathcal{B}} (\Delta E, \mathbf{p}) \mathcal{T}^{\nu \star}_{\mathbf{k},\mathcal{B}} (\Delta E, \mathbf{p}). 
\end{align}

One might wonder why this cross section contains $1 + f_\gamma^\text{eq} (\Delta E)$ even for the inverse process.
To avoid this confusion, we provide an explicit proof for the inverse process starting from Eq.~\eqref{eq:CollisionTermApp}:
\begin{align}
	\text{reverse} &= g_\chi g_{\bar \chi} \sum_\mathcal{B} \int\limits_{\mathbf{k}_\chi,\mathbf{k}_{\bar \chi}} \sigma_\mathcal{B} v_\text{rel} \frac{f_\gamma^\text{eq} (\Delta E)}{1 + f_\gamma^\text{eq} (\Delta E)} f_\mathcal{B} (\mathbf{K} - \mathbf{p}) \nonumber \\
	& =  g_\chi g_{\bar \chi} \sum_\mathcal{B} \int\limits_{\mathbf{k}_\chi,\mathbf{k}_{\bar \chi}} \sigma_\mathcal{B} v_\text{rel}\, e^{ - \frac{\Delta E}{T}} f_\mathcal{B}^\text{eq} (\mathcal{E}_\mathcal{B} + \mathbf{K}^2 / 8 \mu) \times \frac{n_\mathcal{B}}{n_\mathcal{B}^\text{eq}} \nonumber \\
	& =  \sum_\mathcal{B} \Bigg( \frac{g_\chi g_{\bar \chi}}{n_{\chi}^{\text{eq}} n_{\bar{\chi}}^{\text{eq}}} \int\limits_{\mathbf{k}_\chi,\mathbf{k}_{\bar \chi}} e^{- \mathbf{k}_\chi^2 / 4 \mu} e^{- \mathbf{k}_{\bar\chi}^2 / 4 \mu} \sigma_{\mathcal{B}} v_{\text{rel}} \Bigg)  \times \frac{n_{\chi}^{\text{eq}} n_{\bar{\chi}}^{\text{eq}}}{n_{\mathcal{B}}^{\text{eq}}} n_{\mathcal{B}} \nonumber \\
	& =  \sum_\mathcal{B} \langle \sigma_{\mathcal{B}} v_{\text{rel}} \rangle \times \frac{n_{\chi}^{\text{eq}} n_{\bar{\chi}}^{\text{eq}}}{n_{\mathcal{B}}^{\text{eq}}} n_{\mathcal{B}}.
\end{align}
 In the first line we insert the definition of the generalized cross section.
In the second line we use the relation $f_\gamma^\text{eq}(\Delta E) / [1 + f_\gamma^\text{eq} (\Delta E)] = e^{- \Delta E / T}$ and the approximation of kinetic equilibrium $f_\mathcal{B} = f_\mathcal{B}^\text{eq} n_\mathcal{B} / n_\mathcal{B}^\text{eq}$.
In the third line we use the diluteness of the bound state $f_\mathcal{B}^\text{eq} = e^{- (\mathcal{E}_{\mathcal B} + \mathbf{K}^2 / 8 \mu) / T}$ and the energy conservation $e^{- \Delta E / T} e^{- (\mathcal{E}_{\mathcal B} + \mathbf{K}^2 / 8 \mu) / T} = e^{- \mathbf{k}_\chi^2 / 4 \mu} e^{- \mathbf{k}_{\bar\chi}^2 / 4 \mu}$.
This completes the proof of the second term in Eq.~\eqref{eq:gen_Boltz-app}.

\section{Retarded self-energy for massless fermions}
\label{app:retself}

The retarded photon self-energy is defined in terms of greater and lesser self-energies as
\begin{align}
\Pi_{\mu \nu}^R(x-y) &= \theta(x^0-y^0) \left[ \Pi^{-+}_{\mu \nu}(x-y) - \Pi^{+-}_{\mu \nu}(x-y) \right],\label{eq:retardeddef} \\
\Pi^{+-}_{\mu \nu}(x-y)&= g^2 \Tr[ \gamma_{\mu} S^{+-}(x-y) \gamma_{\nu} S^{-+}(y-x)],\\
\Pi^{-+}_{\mu \nu}(x-y)&= g^2 \Tr[ \gamma_{\mu} S^{-+}(x-y) \gamma_{\nu} S^{+-}(y-x)].
\end{align}
where the two-point functions of the fermionic bath-particles are defined as
\begin{align}
S^{-+}_{ij}(x-y)&\equiv\langle \psi_i(x) \bar{\psi}_j(y) \rangle, \\ S^{+-}_{ij}(x-y)&\equiv-\langle\bar{\psi}_j(y)  \psi_i(x) \rangle.
\end{align}
In the free limit and in thermal equilibrium, their Fourier transform is given by
\begin{align}
S^{+-}(K)&= -\slashed{K}(2\pi)\delta(K^2)\left[-\theta(-K^0) + f^{\text{eq}}_{\psi}(|K^0|) \right],\\
S^{-+}(K)&= -\slashed{K}(2\pi)\delta(K^2)\left[-\theta(+K^0) + f^{\text{eq}}_{\psi}(|K^0|) \right].
\end{align}
$f^{\text{eq}}_{\psi}$ is the Fermi-Dirac equilibrium phase-space distribution. In the free and equilibrium limit, the terms needed for the self-energy can be expressed in Fourier space as:
\begin{align}
S^{+-}(x-y)&= \int \frac{\text{d}^4 k}{(2\pi)^4} e^{-ik(x-y)} S^{+-}(k) \nonumber  \\
&= \int \frac{\text{d}^3 k}{(2\pi)^3 2 |\mathbf{k}|} \slashed{K} \left[ - f^{\text{eq}}_{\psi}(|\mathbf{k}|) e^{-ik(x-y)} - (1-f^{\text{eq}}_{\psi}(|\mathbf{k}|))  e^{ik(x-y)} \right] \nonumber \\
&\equiv \int \frac{\text{d}^3 k}{(2\pi)^3 2 |\mathbf{k}|} \slashed{K} \left[ - \alpha_{\mathbf{k}} e^{-ik(x-y)} - \beta_{\mathbf{k}} e^{ik(x-y)} \right], \\
S^{-+}(x-y)&= \int \frac{\text{d}^4 k}{(2\pi)^4} e^{-ik(x-y)} S^{-+}(k) \nonumber \\
&= \int \frac{\text{d}^3 k}{(2\pi)^3 2 |\mathbf{k}|} \slashed{K} \left[ (1-f^{\text{eq}}_{\psi}(|\mathbf{k}|))  e^{-ik(x-y)} + f^{\text{eq}}_{\psi}(|\mathbf{k}|)  e^{ik(x-y)} \right]\nonumber \\
&\equiv\int \frac{\text{d}^3 k}{(2\pi)^3 2 |\mathbf{k}|} \slashed{K} \left[ \beta_{\mathbf{k}}  e^{-ik(x-y)} + \alpha_{\mathbf{k}}  e^{ik(x-y)} \right],\\
\theta(x^0-y^0) &= - \frac{1}{2\pi i} \int \text{d}k \frac{1}{k^0+i\epsilon} e^{-i k^0(x^0-y^0)} .
\end{align}
Where $\alpha$ and $\beta$ are short notation for the phase space density and the Pauli blocking, respectively.

Using these equations, the retarded self-energy can be written in Fourier space as
\begin{align}
\Pi_{\mu \nu}^R(P)&= g^2\int \text{d}^4 z e^{iPz} \int \frac{\text{d}k^0 }{(2\pi)}  e^{-i k^0 z^0}  \int \frac{\text{d}^3 k_1}{(2\pi)^3 2 |\mathbf{k}_1|} \int \frac{\text{d}^3 k_2}{(2\pi)^3 2 |\mathbf{k}_2|} \Tr[\gamma_{\mu} \slashed{K}_1 \gamma_{\nu} \slashed{K}_2] \frac{i}{k^0+i\epsilon}\times \nonumber \\
&\bigg\{ \left(\beta_{\mathbf{k}_1}  e^{-ik_1 z} + \alpha_{\mathbf{k}_1}  e^{ik_1z} \right) \left( - \alpha_{\mathbf{k}_2} e^{ik_2z} - \beta_{\mathbf{k}_2} e^{-ik_2z}\right)- \nonumber \\ &\left( - \alpha_{\mathbf{k}_1} e^{-ik_1z} - \beta_{\mathbf{k}_1} e^{ik_1z}\right)\left(\beta_{\mathbf{k}_2}  e^{ik_2 z} + \alpha_{\mathbf{k}_2}  e^{-ik_2z} \right) \bigg\} \nonumber \\
&= g^2\int \frac{\text{d}^3 k_1}{(2\pi)^3 2 |\mathbf{k}_1|} \int \frac{\text{d}^3 k_2}{(2\pi)^3 2 |\mathbf{k}_2|} \Tr[\gamma_{\mu} \slashed{K}_1 \gamma_{\nu} \slashed{K}_2] (2\pi)^3\times \nonumber  \\
&\bigg\{\alpha_{\mathbf{k}_1} \alpha_{\mathbf{k}_2} \left( \frac{i \delta^3(\mathbf{p}-\mathbf{k}_1-\mathbf{k}_2)}{P^0-|\mathbf{k}_1|-|\mathbf{k}_2| + i \epsilon} -  \frac{i \delta^3(\mathbf{p}+\mathbf{k}_1+\mathbf{k}_2)}{P^0+|\mathbf{k}_1|+|\mathbf{k}_2| + i \epsilon}\right) \nonumber \\ & + \beta_{\mathbf{k}_1} \beta_{\mathbf{k}_2} \left( \frac{i \delta^3(\mathbf{p}+\mathbf{k}_1+\mathbf{k}_2)}{P^0+|\mathbf{k}_1|+|\mathbf{k}_2| + i \epsilon} - \frac{i \delta^3(\mathbf{p}-\mathbf{k}_1-\mathbf{k}_2)}{P^0-|\mathbf{k}_1|-|\mathbf{k}_2| + i \epsilon} \right) \nonumber  \\
&+\alpha_{\mathbf{k}_1} \beta_{\mathbf{k}_2} \left( \frac{i \delta^3(\mathbf{p}-\mathbf{k}_1+\mathbf{k}_2)}{P^0-|\mathbf{k}_1|+|\mathbf{k}_2| + i \epsilon} -  \frac{i \delta^3(\mathbf{p}+\mathbf{k}_1-\mathbf{k}_2)}{P^0+|\mathbf{k}_1|-|\mathbf{k}_2| + i \epsilon}\right) \nonumber \\ &+ \beta_{\mathbf{k}_1} \alpha_{\mathbf{k}_2} \left( \frac{i \delta^3(\mathbf{p}+\mathbf{k}_1-\mathbf{k}_2)}{P^0+|\mathbf{k}_1|-|\mathbf{k}_2| + i \epsilon} - \frac{i \delta^3(\mathbf{p}-\mathbf{k}_1+\mathbf{k}_2)}{P^0-|\mathbf{k}_1|+|\mathbf{k}_2| + i \epsilon} \right) \bigg\} .
\end{align}
After rearrangement, the latter equation can be brought into a convenient form
\begin{align}
&\Pi_{\mu \nu}^R(P)= g^2 \int \frac{\text{d}^3 k_1}{(2\pi)^3 2 |\mathbf{k}_1|} \int \frac{\text{d}^3 k_2}{(2\pi)^3 2 |\mathbf{k}_2|} \Tr[\gamma_{\mu} \slashed{K}_1 \gamma_{\nu} \slashed{K}_2] (2\pi)^3\times \nonumber  \\
&\bigg\{ \left[1-f^{\text{eq}}_{\psi}(|\mathbf{k}_1|) - f^{\text{eq}}_{\psi}(|\mathbf{k}_2|) \right] \left[ \frac{i \delta^3(\mathbf{p}+\mathbf{k}_1+\mathbf{k}_2)}{P^0+|\mathbf{k}_1|+|\mathbf{k}_2| + i \epsilon} - \frac{i \delta^3(\mathbf{p}-\mathbf{k}_1-\mathbf{k}_2)}{P^0-|\mathbf{k}_1|-|\mathbf{k}_2| + i \epsilon} \right] \nonumber \\
&+ \left[ f^{\text{eq}}_{\psi}(|\mathbf{k}_2|) - f^{\text{eq}}_{\psi}(|\mathbf{k}_1|) \right] \left[  \frac{i \delta^3(\mathbf{p}+\mathbf{k}_1-\mathbf{k}_2)}{P^0+|\mathbf{k}_1|-|\mathbf{k}_2| + i \epsilon}- \frac{i \delta^3(\mathbf{p}-\mathbf{k}_1+\mathbf{k}_2)}{P^0-|\mathbf{k}_1|+|\mathbf{k}_2| + i \epsilon} \right] \bigg\},\label{eq:retardedselfenergy}
\end{align}
which is used in section~\ref{sec:nlo} to analyze the various NLO contributions.

\section{Next-to-leading order contributions in more detail}
In this appendix, we share some details of the next-to-leading order computation of the mediator spectral function, as well as some more detailed discussion about the individual contributions.

\subsection{Contour integration for vacuum part}
\label{app:vaccontour}
\begin{figure}[h!]
	\centering
	\includegraphics{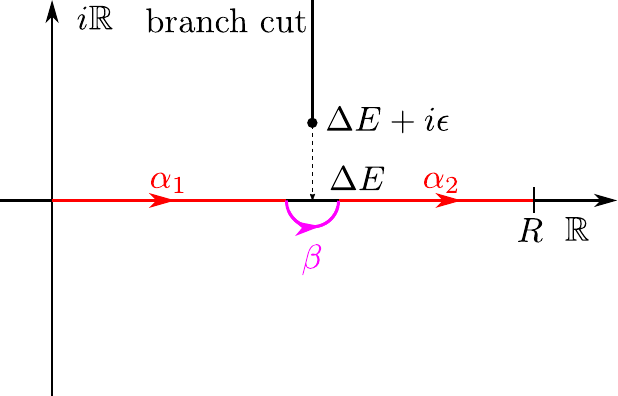}
	\caption{Contour used to compute the vacuum part. Black dot: the pole of the function $G_\epsilon$; $\alpha_1$, $\alpha_2$: paths on the real line; $\beta$: path in the complex plane (half circle) avoiding the singularities of $G_\epsilon$.}
	\label{fig:contour_vac}
\end{figure}

To evaluate the vacuum part of the cross section in Eq.~(\ref{eq:vacxsec}), one has to compute
\begin{align}
	R_0 = \lim_{R\to\infty} \lim_{\epsilon\searrow 0} \frac{1}{\pi} \int_0^R \text{d} |\mathbf{p}| \Im \left[ G_{\epsilon}(|\mathbf{p}|) \right],
\end{align}
for
\begin{align}
	G_{\epsilon}(|\mathbf{p}|) &= \frac{\mathbf{p}^2 (\Delta E^2-\mathbf{p}^2 )\left(\mathbf{p}^2-3\Delta E^2\right)}{3\Delta E^3\left[ (\Delta E+ i \epsilon)^2-\mathbf{p}^2  \right]^2}  \left[\ln\left(\frac{(\Delta E+ i \epsilon)^2-\mathbf{p}^2 }{-\mu^2_0} \right)-\frac{5}{3} \right].
\end{align}
Consider the analytic continuation $|\mathbf{p}| \to z$ and choose the branch cut as shown in fig.~\ref{fig:contour_vac}. In order to avoid integrating thought the singularity at $z=\Delta E+i\epsilon$ we deform the contour and integrate over $\gamma = \alpha_1 + \beta + \alpha_2$ (see fig.~\ref{fig:contour_vac}), where
\begin{align}
	\alpha_1(t) &= t,					&&t \in [0,\Delta E-r],	\\
	\beta(t) &= \Delta E + re^{it},		&&t \in [-\pi,0],	\\
	\alpha_2(t) &= t,					&&t \in [\Delta E+r,R].
\end{align}
As the value of the integral does not change when integrating along $\gamma$ and $G_0$ has no singularities on $\gamma$ we get
\begin{align}
	\lim_{\epsilon\searrow 0} \frac{1}{\pi} \int_0^R \text{d} |\mathbf{p}| \Im \left[ G_{\epsilon}(|\mathbf{p}|) \right]
	= \Im \left[ \lim_{\epsilon\searrow 0} \frac{1}{\pi} \int_\gamma \text{d} z\ G_{\epsilon}(z) \right]
	= \Im \left[ \frac{1}{\pi} \int_\gamma \text{d} z\ G_0(z) \right],
	\label{eq:intG0_vac}
\end{align}
where
\begin{align}
	G_0(z) &= \frac{z^2 \left(z^2-3\Delta E^2\right)}{3\Delta E^3 (\Delta E^2 - z^2)}  \left[\ln\left(\frac{z^2-\Delta E^2}{\mu^2_0} \right)-\frac{5}{3} \right]	\nonumber\\
	&= \frac{1}{3\Delta E} \left( 2 - \frac{z^2}{\Delta E^2} - \frac{\Delta E}{z+\Delta E} + \frac{\Delta E}{z-\Delta E}  \right)  \left[\ln\left(\frac{z^2-\Delta E^2}{\mu^2_0} \right)-\frac{5}{3} \right].
	\label{eq:G0_vac}
\end{align}
Note that $\int_\gamma \text{d} z\ G_0(z)$ does not depend on $r$. Moreover, as $G_0$ is purely real on $\{ x\in\mathbb{R} \mid x>\Delta E \}$ the integration along $\alpha_2$ in Eq.~(\ref{eq:intG0_vac}) gives no contribution. By considering each summand in Eq.~(\ref{eq:G0_vac}) separately and taking the limit $r \searrow 0$ we finally obtain
\begin{align}
	R_0 = \lim_{r\searrow 0} \Im \left[ \frac{1}{\pi} \int_{\alpha_1+\beta} \text{d} z\ G_0(z) \right] = \frac{1}{3} \left[ \ln\left( \frac{\Delta E^2}{\mu^2_0/4} \right) - \frac{10}{3} \right].
\end{align}

\subsection{Scale dependence of the vacuum part}
\label{app:scale}
\begin{figure}[H]
\centering
\includegraphics[scale=0.44]{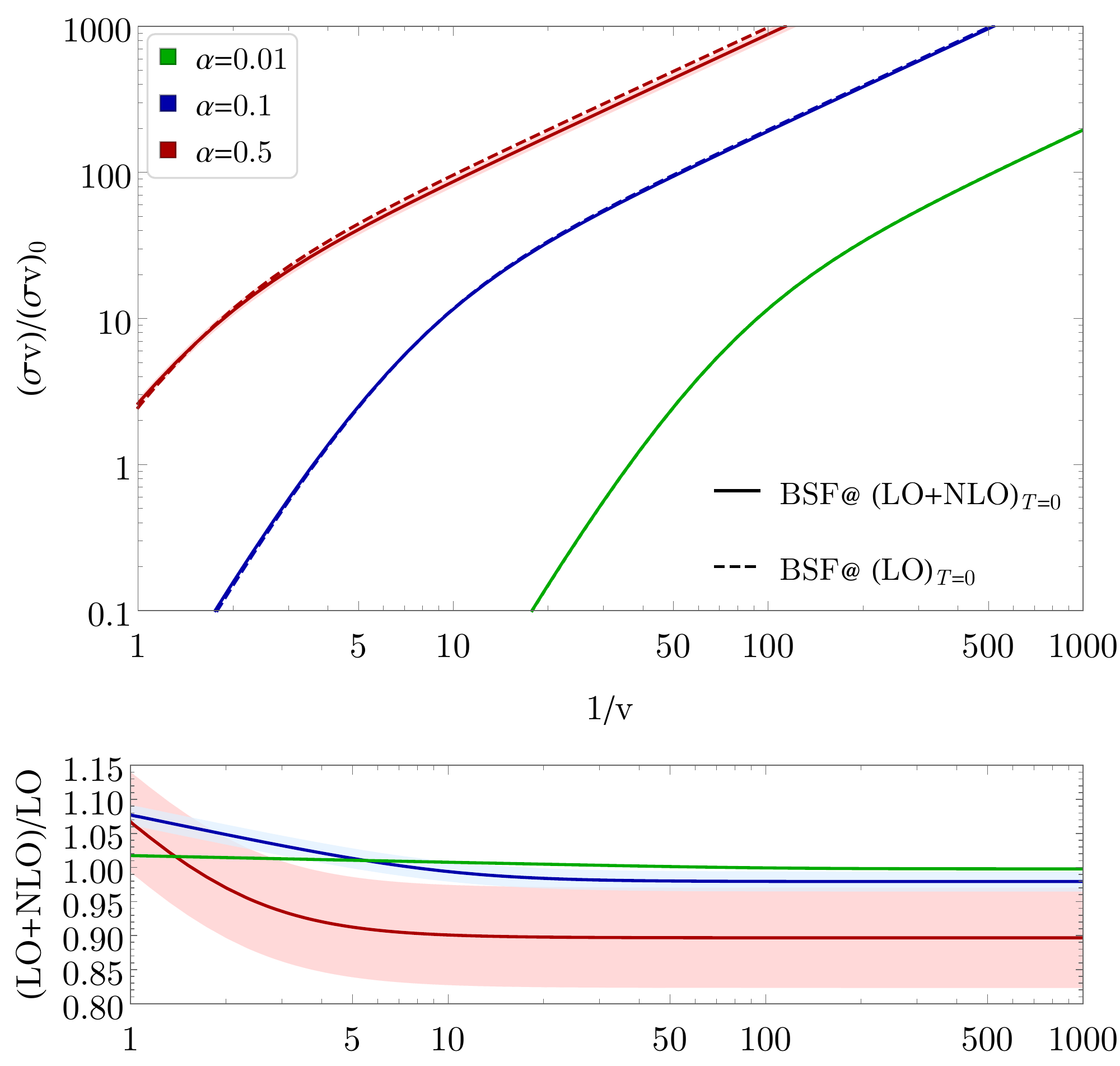}
\caption{\emph{Upper panel: }BSF cross section normalized to the tree-level annihilation cross section at LO (dashed lines) and NLO (solid lines) in dependence of the inverse relative velocity for different coupling strengths $\alpha$ while fixing the renormalization scale to $\mu_0=E_{100}=m_{\chi} \alpha^2/4$ (assuming $N=1$). Shaded areas (barely visible) indicate the scale uncertainty by varying the renormalization scale between $\mu_0=2 E_{100}$ and $\mu_0=E_{100}/2$. \emph{Lower panel: } Relative correction of the NLO contribution with respect to the LO cross section with the renormalization scale fixed at $\mu_0=E_{100}=m_{\chi} \alpha^2/4$ (solid lines). The shaded area indicates the renormalization uncertainty as in the upper panel.}
\label{fig:vac}
\end{figure} 

As discussed in section~\ref{sec:cumputation}, in particular around Eq.~(\ref{eq:vacxsec}), the NLO contribution at zero temperature can be factorized in terms of the LO contribution,
\begin{align}
(\sigma_{100}^{\text{LO}} v_{\text{rel}}) + (\sigma^{\text{NLO}}_{100} v_{\text{rel}})_{T=0}  = (\sigma_{100}^{\text{LO}} v_{\text{rel}}) \bigg\{1+ N\times \frac{\alpha}{3\pi} \left[ \ln\left( \frac{\Delta E^2}{\mu^2_0/4} \right) - \frac{10}{3} \right] \bigg\} \label{eq:vacuumscale} \,,
\end{align}
where $N$ specifies the number of particle species in the thermal plasma and thus also the multiplicity of particles that are running in the loop in the NLO contribution to the on-shell emission. As expected from the renormalization procedure of the occurring UV divergence, the final cross section will feature a renormalization scale dependence $\mu_0$ that needs to be fixed by a physical motivated scale. As the characteristic scale of the BSF process lies around the binding energy, we choose accordingly $\mu_0=E_{100}=m_{\chi} \alpha^2/4$.

In fig.~\ref{fig:vac}, we show the vacuum contribution at LO and NLO in dependence of the inverse relative velocity $v_{\mathrm{rel}}$ for coupling strengths up to the order of the unitarity bound for s-wave SE annihilation, $\alpha = 0.5$ \cite{Smirnov:2019ngs}. In order to estimate the uncertainty that arises from this particular choice of scale, we vary the latter between $\mu_0=2 E_{100}$ and $\mu_0=E_{100}/2$ (shaded area). The upper panel features the general expected behaviour: With larger velocity the relative cross section decreases as BSF is suppressed due to the small overlap of the wave functions. Moreover, larger couplings $\alpha$ lead to a relatively larger BSF cross section. From Eq.~\ref{eq:vacuumscale}, we can infer that for larger couplings the NLO correction becomes more relevant and hence also the dependence of the scale uncertainty, as visible in the lower panel. Note that we have taken the conservative choice of $N=1$. For larger $N$ not only the correction itself, but also the scale uncertainty is expected to be more pronounced.

For the renormalization of the occurring UV divergence, we had chosen the simplest choice for the renormalization scheme, namely the $\overline{\text{MS}}$ scheme. However, a more physical motivated approach would be to use an on-shell scheme for the renormalization of the photon propagator since the photon is on-shell in the corresponding interference terms. However, this unnecessarily complicates the cancellation of the infrared divergences. While it would be academically interesting to study the application of the on-shell scheme and to compare it with the $\overline{\text{MS}}$ scheme in more detail, we leave this investigation for future work.

\subsection{Contour integration for the finite temperature part}
\label{app:contourFT}

The function $G^{\sigma_1\sigma_2}_\epsilon(z) = G^{\sigma_1\sigma_2}_\epsilon(z,\tau,|\mathbf{k}|)$ in the finite temperature part of the cross section in Eq.~(\ref{eq:ftxsec}) has three types of poles:
\begin{itemize}
	\item the double poles of the squared mediator propagator at $\pm (\Delta E + i\epsilon)$,
	\item the single poles of the retarded self-energy at $w_{p/m}$, $w_p$ lying in the upper half and $w_m$ lying in the lower half of the complex plane,
	\item and the poles of $F^{\sigma_1\sigma_2}$, which do not lie on the real line.
\end{itemize}
When $\epsilon$ goes to 0, $w_0 \equiv \Delta E+i\epsilon$ goes to $z_0=\Delta E$ which lies on the positive real line. $w_p$ and $w_m$ go to
\begin{align}
	z_{p/m} &= \lim_{\epsilon\searrow 0} w_{p/m} \equiv -\tau \sigma_1|\mathbf{k}| \pm \mathrm{sign}(\Delta E + \sigma_1|\mathbf{k}|)\sqrt{\tau^2|\mathbf{k}|^2+\Delta E^2+2\sigma_1\Delta E|\mathbf{k}|},
\end{align}
which can lie on the positive real line, as well.

\begin{figure}[h!]
	\centering
	\includegraphics{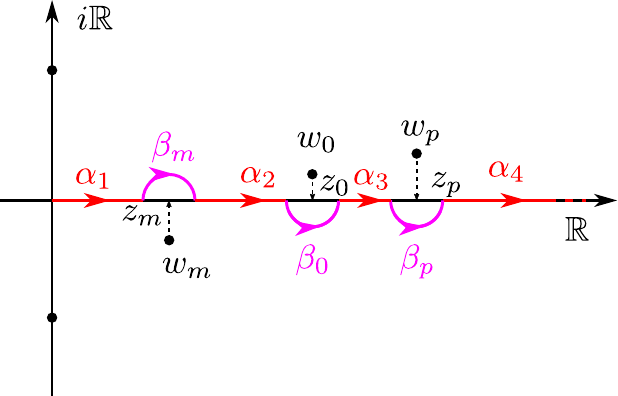}
	\caption{Contour used to compute the finite temperature part. Black dots: poles of the function $G^{\sigma_1\sigma_2}_\epsilon$; $z_0$, $z_p$, $z_m$: location of the poles of $G^{\sigma_1\sigma_2}_0$ which lie on the real line; $w_0$, $w_p$, $w_m$: location of the poles of $G^{\sigma_1\sigma_2}_\epsilon$ which go to $z_0$, $z_p$, $z_m$ as $\epsilon\searrow 0$; $\alpha_1$, $\alpha_2$, $\alpha_3$, $\alpha_4$: paths on the real line; $\beta_0$, $\beta_p$, $\beta_m$: paths in the complex plane (half circles) avoiding the singularities of $G^{\sigma_1\sigma_2}_\epsilon$.}
	\label{fig:contour_n}
\end{figure}

For the moment let us assume that $z_{p/m}$ are real and positive. To avoid integrating through the singularities we deform the path of integration as shown in fig.~\ref{fig:contour_n}. 
The deformation of the path does not affect the value of the integral over $|\mathbf{p}|$ as $F^{\sigma_1\sigma_2}$ is holomorphic in some neighborhood of the real line. Since $G^{\sigma_1\sigma_2}_0$ has no poles on the new path $\gamma$, we get
\begin{align}
	\lim_{\epsilon \searrow 0} \int_0^\infty \text{d}|\mathbf{p}|\ G^{\sigma_1\sigma_2}_\epsilon(|\mathbf{p}|,\tau,|\mathbf{k}|)
	&= \lim_{\epsilon \searrow 0} \int_\gamma \text{d} z\ G^{\sigma_1\sigma_2}_\epsilon(z,\tau,|\mathbf{k}|)
	= \int_\gamma \text{d} z\ G^{\sigma_1\sigma_2}_0(z,\tau,|\mathbf{k}|).
\end{align}

As $G^{\sigma_1\sigma_2}_0(z)$ is real valued on the real line, the integration along $\alpha_1,\dots,\alpha_4$ only gives a real contribution.
Moreover, it follows that the Laurent coefficients of $G^{\sigma_1\sigma_2}_0$ at $z=z_0$, $z=z_p$, and $z=z_m$, respectively, are real. Using Laurent series expansion it is straightforward to verify that
\begin{align}
	\Im\left( \int_{\beta_0} G^{\sigma_1\sigma_2}_0(z) \text{d}z \right) &= \pi \Res(G^{\sigma_1\sigma_2}_0,z_0),	\\
	\Im\left( \int_{\beta_{p/m}} G^{\sigma_1\sigma_2}_0(z) \text{d}z \right) &= \pm \pi \Res(G^{\sigma_1\sigma_2}_0,z_{p/m}).
\end{align}

Finally we get
\begin{align}
	&\lim_{\epsilon\searrow 0} \int_0^{\infty} \text{d}|\mathbf{p}| \; \Im\big[G^{\sigma_1\sigma_2}_{\epsilon}(|\mathbf{p}|,\tau,|\mathbf{k}|)\big] 
	= \lim_{\epsilon\searrow 0} \int_0^{\infty} \text{d}z \; \Im\big[G^{\sigma_1\sigma_2}_{\epsilon}(z,\tau,|\mathbf{k}|)\big]												\nonumber\\
	&= \pi \cdot \Big[ \Res(G^{\sigma_1\sigma_2}_0,z_0) + \Res(G^{\sigma_1\sigma_2}_0,z_p) - \Res(G^{\sigma_1\sigma_2}_0,z_m) \Big], \text{ for } z_{p/m}\in\mathbb{R}^+.	\label{eq:sumres}
\end{align}

If one of the poles $z_{p/m}$ is negative or not real then the corresponding term in Eq.~(\ref{eq:sumres}) vanishes as there is no need to deform the path of integration. In particular, by the same argument the poles on the imaginary axis, originating from the equilibrium phase-space distributions in  Eq.~(\ref{eq:NLO}), do no contribute. 

\subsection{Individual contributions of the finite temperature part}
\label{app:indFT}

\begin{figure}[H]
\centering
\includegraphics[scale=0.55]{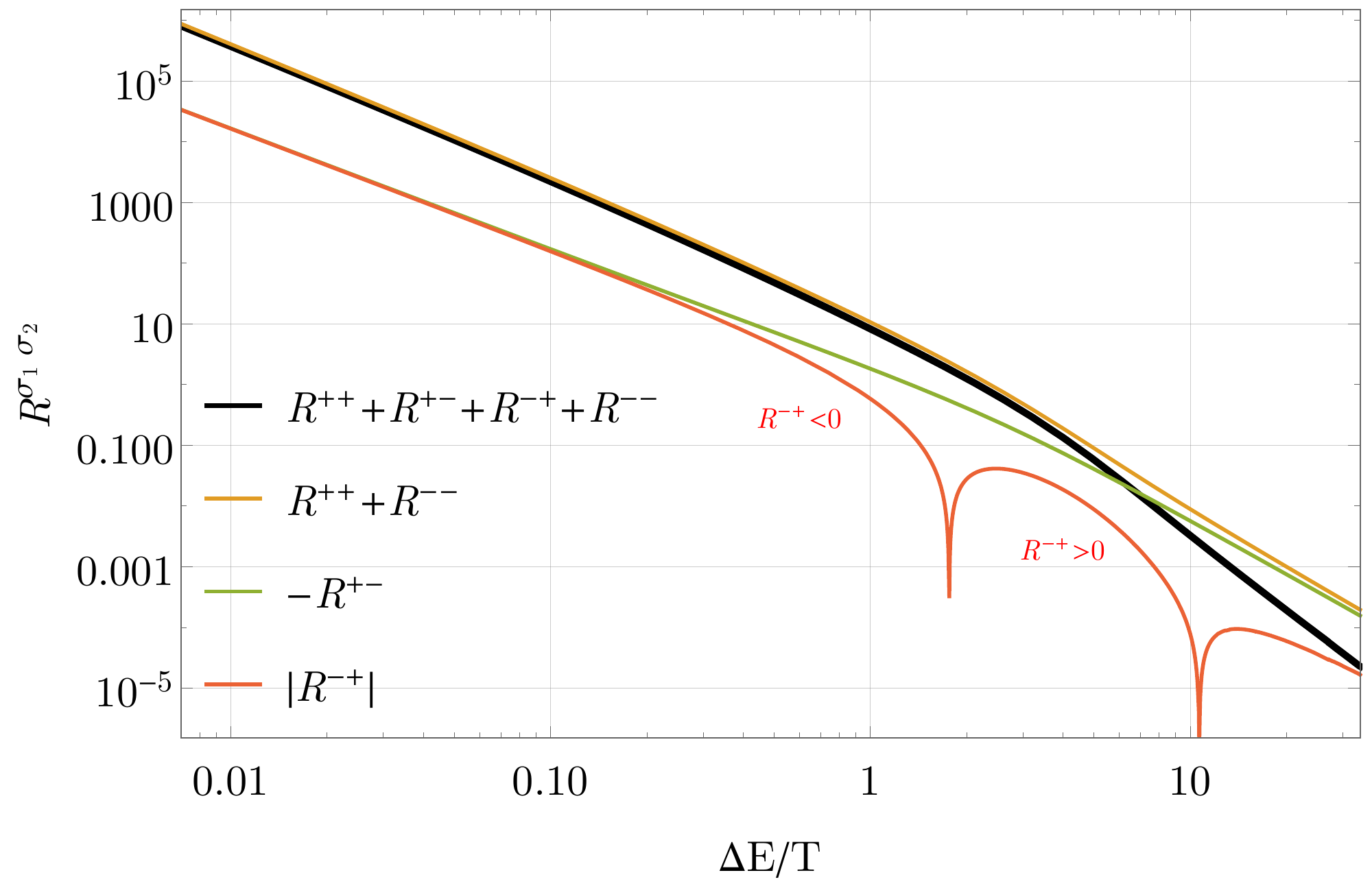}
\caption{The plot shows the individual $R^{\sigma_1 \sigma_2}$ functions and their dependence on $\Delta E/T$.}
\label{fig:rfunction}
\end{figure}

The $R^{\sigma_1 \sigma_2}$ functions, in Eq.~(\ref{eq:Rpp}) to Eq.~(\ref{eq:Rmm}), are all dimensionless. Therefore they can only dependent on the ratio $\Delta E/T$ for massless mediators. This fact makes the relic abundance computation efficient, since one can tabulate the sum over all these functions in advance depending on a single variable. 

For a selected range we show the individual contributions in fig.~\ref{fig:rfunction}. $R^{++}$ and $R^{--}$ contain the BSF via bath-particle and anti-particle scattering and are the most dominant contributions. The double pole in $R^{+-}$ contributes negatively, and $R^{-+}$ has some oscillating features. The latter contains the finite temperture part of the off-shell mediator decay into bath particles as well as other single and double pole residues. Depending on which part dominates inside $R^{-+}$, the function takes positive and negative values. 

Most important is the value of the sum over all functions at $T/\Delta E \sim 1$, since there one would roughly enter the exponential depletion phase of the relic abundance if ionization equilibrium is maintained. One can see that for a very precise determination of the relic abundance, the inclusion of all terms are needed at least in this case. Throughout this work we take all of them into account.

\bibliographystyle{JHEP}
\bibliography{BSF}








\end{document}